\documentclass[11pt,a4paper]{article}

\usepackage{caption,subcaption}
\usepackage{multirow}
\usepackage{amsmath}
\usepackage{amsfonts}

\usepackage{amssymb}
\usepackage{hyperref}
\usepackage{algorithm}
\usepackage{algpseudocode}
\usepackage{graphicx}
\usepackage{authblk}
\date{}
\AtBeginDocument{%
  \providecommand\BibTeX{{%
    \normalfont B\kern-0.5em{\scshape i\kern-0.25em b}\kern-0.8em\TeX}}}

\begin{document}

\title{Deep Learning to Jointly Schema Match, Impute, and Transform Databases}


\author[1]{Sandhya Tripathi \thanks{sandhyat@wustl.edu}}
\author[1]{Bradley A. Fritz \thanks{bafritz@wustl.edu}}
\author[2]{Mohamed Abdelhack\thanks{mohamed.abdelhack.37a@kyoto-u.jp}}
\author[1]{Michael S. Avidan \thanks{avidanm@wustl.edu}}
\author[3]{Yixin Chen \thanks{ychen25@wustl.edu}}
\author[1]{Christopher R. King \thanks{christopherking@wustl.edu}}
\affil[1]{Department of Anesthesiology, Washington
               University in St Louis, MO, USA}
\affil[2]{Krembil Centre for Neuroinformatics, Centre for Addiction and Mental Health, Toronto, Ontario, Canada}
\affil[3]{Department of Computer Science and
               Engineering, Washington University in St Louis, St Louis,
               MO, USA}

% \author{Sandhya Tripathi}
% % \authornote{Department of Anesthesiology, Washington
% %               University in St Louis, MO, USA}
% % \email{sandhyat@wustl.edu}
% \author{Bradley A. Fritz}
% % \email{bafritz@wustl.edu}

% \author{Mohamed Abdelhack}
% % \authornote{Krembil Centre for Neuroinformatics, Centre for Addiction and Mental Health, Toronto, Ontario, Canada}
% % \email{mohamed.abdelhack.37a@kyoto-u.jp}

% \author{Michael S. Avidan}
% % \email{avidanm@wustl.edu}

% \author{Yixin Chen}
% % \authornote{Department of Computer Science and
% %               Engineering, Washington University in St Louis, St Louis,
% %               MO, USA}
% % \email{ychen25@wustl.edu}

% \author{Christopher R. King}
% % \email{christopherking@wustl.edu}

% \renewcommand{\shortauthors}{..}


\maketitle

\begin{abstract}
An applied problem facing all areas of data science is harmonizing data sources.
Joining data from multiple origins with unmapped and only partially overlapping features is a prerequisite to developing and testing robust, generalizable algorithms, especially in health care.
We approach this issue in the common but difficult case of numeric features such as nearly Gaussian and binary features, where unit changes and variable shift make simple matching of univariate summaries unsuccessful.
We develop two novel procedures to address this problem.
First, we demonstrate multiple methods of ``fingerprinting'' a feature based on its associations to other features.
In the setting of even modest prior information, this allows most shared features to be accurately identified.
Second, we demonstrate a deep learning algorithm for translation between databases.
Unlike prior approaches, our algorithm takes advantage of discovered mappings while identifying surrogates for unshared features and learning transformations.
In synthetic and real-world experiments using two electronic health record databases, our algorithms outperform existing baselines for matching variable sets, while jointly learning to impute unshared or transformed variables.
\end{abstract}


%   <concept>
%       <concept_id>10010405.10010444.10010447</concept_id>
%       <concept_desc>Applied computing~Health care information systems</concept_desc>
%       <concept_significance>500</concept_significance>
%       </concept>
%   <concept>
%       <concept_id>10010147.10010257.10010293.10010294</concept_id>
%       <concept_desc>Computing methodologies~Neural networks</concept_desc>
%       <concept_significance>500</concept_significance>
%       </concept>
%  </ccs2012>

% \ccsdesc[500]{Applied computing~Health care information systems}
% \ccsdesc[500]{Computing methodologies~Neural networks}


\vspace{1cm}

\section{Introduction}

A perennial problem in modern data science is integrating multiple data sources.
We focus on the problem in which similar or identical concepts are measured in distinct data sources, commonly referred to as the schema matching problem \cite{dhamankar_imap_2004}.
Manually reconciling large scale databases where the ``columns'' (database attributes) have not been previously mapped to a common ontology is an expensive and error prone process.
Our motivating example comes from healthcare, where electronic health records (EHR) store thousands of distinct types of observations, many of them using redundant or ambiguous names.
Combining these databases is a necessity for the development of large-scale registries, generalizable clinical decision support tools, and measuring performance in small subpopulations.

Where possible, matching the meta-data from two sources (such as column names) is a useful approach, reviewed briefly below.
However, in many cases it is not sufficient and using summary statistics of a column as a fingerprint (instance-based matching) is both necessary and powerful \cite{yang_effective_2008}.
The EHR context is particularly challenging for instance-based matching using summary statistics for several reasons.
First, a large fraction of EHR data columns are binary, meaning that other than the frequency and missingness ratio no univariate summary statistics are possible.
All binary columns with similar frequencies are indistinguishable potential matches in a univariate approach.
Second, a shift in variable distributions is the norm when concatenating databases from distinct contexts.
This can occur both naturally, for example, when populations in two hospitals have different weight and age distributions, and artificially when systems use incompatible units to record observations. 
For example, in EHR data originating in the United States, weight, height, pressure, and concentration could be recorded in SI or US conventional units.
%A closely related problem occurs when identical or very similar medications with distinct trade names are listed with different concentrations or uninformative units like ``tablet''.
More generally, up to a unit change, any approximately normally distributed set of variables will be possible matches.
Finally, some columns may be simple but nonlinear re-parameterizations of others; for example, body surface area of height and weight.
This final problem confounds many prior approaches to matching the distribution of features from multiple databases.

% \textbf{Contribution} 
We propose a novel deep-learning based solution to the schema matching problem inspired by distributional semantics.
That is, the relationships between entries for a given column and other database columns define the semantics of that column.
For example, indicators for a patient having diabetes, overweight, high blood pressure, and hyperlipidemia are a cluster known clinically as ``metabolic syndrome,"  and this cluster of correlation could be identified in each dataset. 
Relationships between variables are plausibly unchanged across databases even with substantial variable shift, especially if they are causal relationships.
%We discuss special cases where those semantics can be unique defined and matched across databases without other prior knowledge, but our main contribution is the case where some prior knowledge connecting the two databases exists.
In some special cases, meaningful matching can be performed with no prior information, but we focus on the more realistic case where a few columns in two databases are known to be semantically identical.
That is, we assume that at least some features are mapped by auxiliary knowledge or manual verification (known-mapped columns).
We then use autoencoders to compress the data to a relatively low-dimension latent space, using the known-mapped columns to anchor the latent spaces of the autoencoders together. 
Those autoencoders contain the pattern of dependency between columns (the semantics) in each database.

\begin{figure*}[h!]
    % \centering
    % \begin{subfigure}[b]{0.478\textwidth}
        \centering
        \includegraphics[width=1\textwidth]{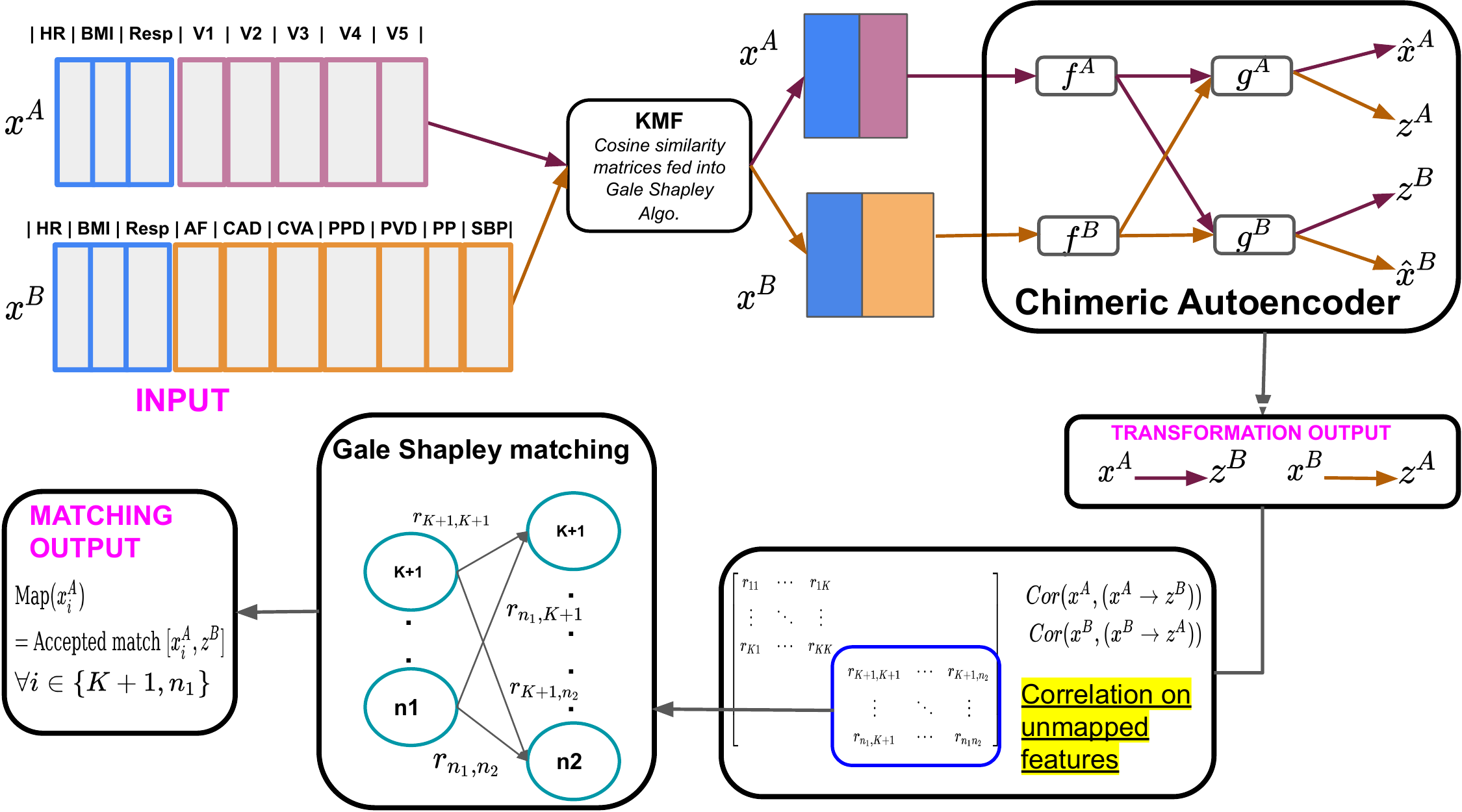}
        \caption{An illustration of the KMF initialized chimeric schema matching algorithm with module-wise description at the start of Section \ref{sec: proposed_method}. It also depicts how the input datasets could be differently sized with few mapped features (first three blue coloured columns with common feature names) and majority unmapped (pink and orange coloured columns with ambiguous feature names).}
        % {}    
        \label{fig: Schema_full_2stage_Chimeric}
\end{figure*}

To extract human-readable interpretations from the semantics of the autoencoders, we take a novel knowledge distillation approach.
We create ``chimeric'' encoders that transform from the format of the first database to the second.
To identify additional matching variables, we then compute cross-format associations between the original and transformed data.
We use a simple matching algorithm to pair columns from the first database to the second, maximizing the sum of correlations between proposed matches.
A step-down procedure is used to reject low correlation proposed matches while controlling the overall false discovery rate.
The discovery of low-dimensional transformations is facilitated by graphical analysis and mutual information between the original and transformed data.
These chimeric encoders also function to impute features between databases where an exact match is not found.
This is a very common occurrence in medical databases where a feature is not measured in some sources, but strong correlates are.
For example, in our real dataset, the diagnosis of atrial fibrillation (a common abnormal heart rhythm) may not be explicitly noted, but the presence of medications related to atrial fibrillation allow it to be inferred reasonably well.
Remarkably, the algorithm does not require the ``imputed'' variable to be present in both databases.
A complete pipeline of our approach is presented in Figure \ref{fig: Schema_full_2stage_Chimeric}.
To summarize, we make the following \textbf{contributions}:
\begin{itemize}
    \item We propose a novel summary statistic (fingerprint) method that solves the schema matching problem with moderate accuracy and high efficiency.
    \item Our algorithm identifies surrogates for features that are not shared by both databases but have a closely-related concept available.
    \item Our chimeric encoders learn to jointly impute and transform samples from one database to another when correlates or re-parameterizations of unshared features are present.
\end{itemize}
%We present a second method which uses a single autoencoder and learns a permutation matrix which optimizes that autoencoder on alternative databases.

\paragraph{Terminology.}
The literature on related problems has used a confusing variety of terms.
Each data source will be represented by a \textit{database} which is a combination of a \textit{dataset} (a set of tuples each of which represents the information on a single object e.g. a patient) and schema information, meta-data, relationships, and restrictions.
Datasets are also organized by \textit{features} (also commonly called \textit{attributes}), which in the simplest cases correspond to a column index in a data matrix.
We will occasionally refer to a \textit{column} as the slice of all data for a feature in a dataset.
Each feature is a measurement of a real-world \textit{variable}, and features in multiple databases can represent the same variable.
We refer to these features as \textit{mapping} to one another.
\textit{Meta-data} is information about a feature, such as its name, data type, and ontology tags.
A \textit{fingerprint} of a feature is a set of summary statistics about that column; this is also commonly called an \textit{embedding}.
The process of discovering features across databases that reflect the same variable is \textit{schema matching}; features that share a variable are \textit{maps} of one another.
When two databases have exactly the same set of variables, they have a \textit{bijective} map or are permutations of one another.
When the variables in one database are a proper subset of another, they have an \textit{onto} map.
When the two share some but not all variables, they have a \textit{partial} map.
% Two unrelated databases have a \textit{null} map. We do not further consider this uninteresting case.

\textbf{Organization.} In Section \ref{sec: related_work}, we start with a brief review of related literature in schema matching and autoencoders. 
In Section \ref{sec: Motivating_example}, we provide details about our motivating example involving EHRs. 
We present the proposed schema matching algorithm in Section \ref{sec: proposed_method}. 
In Section \ref{sec: experiment_results}, we compare the performance of proposed algorithms to baselines and simple alternatives on multiple synthetic and real-data examples.
Finally, in Section \ref{sec: discussion} we discuss the results, alternatives, and agenda for future work.

\subsection{Related work} \label{sec: related_work}
Prior approaches to schema matching fall into a few categories; a recent review is provided in \cite{alwan2017survey}.
First, meta-data about each feature (such as names) can be compared across databases.
Modern deep-learning approaches to language embedding can be used to combine text descriptions and other meta-data of a feature before sequence-pair classification, as in \cite{li2020deepvldb2021}. 
Properties of the schema (such as restrictions) can narrow the set of potential mappings and fit within this category.
Second, if specific examples are present in multiple databases (e.g. shared patients across hospitals), these can be used to match schema using entity matching algorithms \cite{kopcke2010frameworksEM_review}. 
Third, fingerprints of each feature can be compared across databases to find the best matches (instance-based matching).
Authors in \cite{koutras2021valentine} provide a comprehensive review of the above categories with pros and cons of the algorithms available in each category and present an interface that can evaluate different types of schema matching methods on a common metric. 
Practical implementations combine these approaches to complement each other \cite{madhavan_generic_2001}. 

The use of meta data for schema matching is the first approach to be used if the schema has clear and comprehensible attribute names. 
Authors in \cite{bulygin2018combining} review many of the element-level matchers that use the name and description of a column.
They further demonstrate that a hybrid of these matchers (combining different similarity measures between the entity pairs) using machine learning outperforms the individual matchers. 
Recent work by \cite{satti2021unsupervisedHealthcareSchema} uses RoBERTa, a transformer-based deep learning model to obtain the embeddings for the attribute names from a given schema and then computes similarity on the generated embeddings. Another attention-based deep learning solution is proposed by \cite{zhang2021smat} that only uses attribute names and descriptions. 
% \cite{de2018machineIJCAIontomapping} address a general problem of creating a schematic model for a data source, by using some of the existing schematic models from the same domain as training samples. 
% These training samples are embedded attribute names used to train a machine learning model and then optimize a constraint program formulation to obtain a schema. 

Instance-based matching requires a set of summaries of a data column (a fingerprint);
one of many matching algorithms is then used to select the best mapping of features of multiple databases based on the similarity of fingerprints.
Much of this literature uses string data types, where formatting patterns, word frequencies, and neural embedding can identify the semantics of a feature.
For quantitative data, simple characteristics such as the mean, variance, mode, and fraction missing are commonly used \cite{yang_effective_2008,sahay_schema_2020}.
Others have proposed more complex fingerprints for continuous data. 
Dhamankar and colleagues \cite{dhamankar_imap_2004} computed the Kullback-Leibler divergence between each pair of quantitative columns (that is, using the distribution function as a high dimensional summary). 
Jaiswal and colleagues \cite{jaiswal_schema_2013} similarly fingerprint a feature by fitting a Gaussian mixture model to its distribution and compared the fit pairwise to candidate variables.
Mueller and Smola \cite{mueller_recognizing_2019} used a corpus of labeled correspondences between datasets to learn a neural-network based embedding of distributions.
Because of the large (infinite) dimension of possible embedding of continuous columns, variable reduction has been proposed using penalized regression \cite{berlin_database_2002} and random projections \cite{bourennani_content-based_2019}.

Notably, the above proposed fingerprints are functions of a single column.
However, there is good reason to believe that \textit{associations} between variables (and hence features) will tend to be more reproducible across databases.
For example, although the fraction of men at a given hospital can vary, the relationship between sex and height is likely very similar.
Several authors have proposed using the dependencies between columns to define the fingerprint.
Kang and colleagues \cite{kang_schema_2003} proposed a 3 step process: first, in both datasets compute the mutual information of every pair of columns; second, filter potential pairs from across databases as potential mappings based on univariate entropy (or other summary statistics); third, search over assignments from the second database to the first maximizing the similarity of the mutual information matrices. 
They later compared optimization criteria and exhaustive versus heuristic methods for searching over mapping assignments in \cite{kang_schema_2008}. 
Cruz and colleagues \cite{cruz2007privacy} (details unpublished) combined the entropy and mutual information characteristics with a privacy preserving set intersection method to replace graph alignment.
Authors in \cite{zhang_automatic_2011} presented a similar proposal that replaced mutual information with the earth movers distance and took a stepwise approach to correlation clustering and matching the similarity of clusters. 
Rabinovich and Last \cite{rabinovich_scalable_2014} modified the Kang approach by filtering potential assignments on absolute correlation (or an equivalent measure for binary variables) and added noise (or knock-off) columns to force the two databases to have the same number of columns.
In Kang and related approaches, any features which are known to map are fixed during the initialization of the graph search.

Translating between two databases (i.e. estimating for a given row in database A, a row which approximately matches the semantics of database B) is a similar problem, which has been approached by one prior method.
RadialGAN \cite{yoon_radialgan_2018} attempts to generate samples from auxiliary databases in the format of a primary database to support training a classifier.
Briefly, for each database an autoencoder (AE) is fitted.
To encourage the latent spaces to have the same semantics, two modification to a standard AE are taken.
First a \textit{cycle consistency} loss is added while training the autoencoders by comparing encoded examples to re-encoded translated examples (Mapping A to B, then back to A).
Second, a discriminator is trained to distinguish true samples from generated samples, and the loss of that discriminator maximized, similar to a GAN except using the latent representation of alternative databases to supply the generator (the decoder) rather than noise.
After training (when the discriminator can no longer differentiate generated from authentic samples), the generated examples are concatenated to the primary dataset for training a classifier.
The RadialGAN method does not propose to match the schemas directly, but it produces mechanically similar translations to our chimeric encoders and can be used as input to matching methods.

The process of transforming estimated similarities between columns into proposed matches and evaluating the result can itself be a complicated process.
Shraga, Gal, and Roitman review current approaches to this step of the algorithm, including the use of neural networks to learn transformation which optimally adjust estimated similarities \cite{shraga2020adnev}.
Another approach that improves the matches is proposed by \cite{zhang2018reducingUncerCCQ} where algorithms generate questions for crowd-sourced workers to maximize reduction in uncertainty of the matches.
Authors in \cite{yousfi2020towards} also consider this issue while matching multiple schema by finding pairwise schema overlap using some existing similarity measure based schema matches.
%CNNs and RNNs are used by   to propose a variant of second line matcher (2LM). 
%The main idea is to adjust and evaluate the similarity matrices generated by any arbitrary matcher where the adjustors use CNNs to solve a binary classification problem. 
%Our chimeric encoders contribute towards generating the similarity matrices whereas the setup by \cite{shraga2020adnev} assumes that these similarity matrices exist, hence our usecases do not overlap.

These approaches have substantial deficiencies in our motivating example.
For example, metadata may be absent or misleading.  
The most common and useful metadata is often the feature name.
In the Epic EHR, many data elements (``smartdata'') have short, meaningless names assigned differently at different sites; some examples are available in Table \ref{tab:smart_data_example} in Appendix \ref{app: real_Data_details}.
Where present, names that appear similar may be distinct concepts, such as ``Ur Creatinine'' and ``POC Creatinine'' or may contain ambiguous abbreviations such as ``PHTN'' representing ``pulmonary hypertension,'' ``portal hypertension,'' or ``pre-hypertension'' which are themselves distinct from ``HTN'' or ``hypertension'' as a concept.
%``PT'' indicating ``physical therapy,'' ``patient,'' or ``prothrombin time.''
True database schema properties may not be available or preserved, as data is often exported to a flat format.
Finally, in the healthcare setting, few identical records from the same source will exist between databases in most cases.
That is, from two hospitals in different states or time periods we expect few contemporaneous records of the same patient.
However, we fully expect these methods to produce some successful high-confidence matches, and regard them as one source of known-mapped columns in our approach. 
As mentioned above, instance-based fingerprints are also very limited by binary variables and distributional shifts between data sources.
For example, indicator variables for two diseases with roughly the same frequency would have very similar fingerprints and be vulnerable to mis-mapping.

Generative adversarial networks (GANs) can fulfil some of the same application needs as schema matching.
For example, realistic synthetic data in the format of a target database can augment training a classifier.
Our method uses autoencoders (AE), and many researchers have adapted GANs to improve the training of AE or enforce desired properties; for example, \cite{makhzani2015adversarial} use a discriminator network to make the encoder output similar in distribution to spherical Gaussian data.
% Adversarial autoencoders \cite{makhzani2015adversarial} attempt to improve GAN output by using the hidden code of the autoencoders. 
%Adversarial autoencoders \cite{makhzani2015adversarial} have an additional discriminator network to differentiate between the latent representation (output of AEs' encoder) of real data points and Gaussian noise (when fed into the AEs' encoder). %
After jointly training the AE and discriminator network, Gaussian random variables can be fed into the AE's decoder to generate new data points.
Variational AEs such as $\alpha$-GAN \cite{rosca2017variational} can be used to generate synthetic data with additional distributional requirements. 
However, a discriminator network may not be necessary for these models. The Adversarial Generator-Encoder Networks (AGE) \cite{ulyanov2018takes} model uses the divergence of two induced latent distributions as a kind of pre-made discriminator. 
Expanding this approach with multiple discriminators and generators/encoders such as done in \cite{lazarou2020autoencoding} can stabilize the mapping of data points to a latent space and back, perhaps reducing the mode collapse and training difficulty of many GANs. 
Even though the above methods have the potential to generate augmenting data, they do not directly address all the use cases of a schema-matched dataset and do not exploit the relationships between features across databases.

\section{Methods} \label{sec: methods}

\subsection{Description of motivating example} \label{sec: Motivating_example}
\paragraph{ACTFAST}
Clinicians and informaticists in the Anesthesiology department at Washington University School of Medicine developed a database of surgical patients including preoperative clinical characteristics, laboratory data, high-frequency intraoperative monitor data, and postoperative outcomes.
This database was developed by manually linking multiple existing sources for applications in clinical epidemiology \cite{king_association_2020} and clinical decision support tools \cite{fritz_deep-learning_2019}.
It was also linked to an existing ontology as part of a large registry project \cite{colquhoun_considerations_2020}.
However, in 2018 the hospital switched from the prior combination of EHRs (MetaVision, Allscripts) to Epic.
Although many variables were able to be mapped using manual exploration, ontology links, and hierarchical representations within Epic, others were not.
A data query revealed thousands of potential mapping variables, making exhaustive manual review impossible.
% One challenging set of features was laboratory values, where in both eras ambiguous names and abbreviations (as mentioned in the introduction) forced manual mapping of both to a common ontology (LOINC). 
% TODO: add this back if I get the example working

\paragraph{MIMIC} 
The Medical Information Mart for Intensive Care (MIMIC) dataset includes routinely collected EHR and administrative data on a large number of critically ill patients, and is the most widely used dataset for experimentation with machine learning in critical care \cite{johnson2016mimic}.
One interesting feature of MIMIC is that during the study, the Carevue EHR was replaced with the Metavision EHR.
Data from many systems which indirectly feed into the EHR, such as laboratory values, has a consistent representation over the entire study, but discrete data which is manually charted (the chartevents table) have inconsistent identification numbers (called d\_items) after the change.
Prior projects \footnote{1) \url{https://github.com/USC-Melady/Benchmarking_DL_MIMICIII/blob/master/Codes/mimic3_mvcv/10_get_99plus-features-raw.ipynb} \\ 2) \url{https://github.com/YerevaNN/mimic3-benchmarks/blob/master/mimic3benchmark/resources/itemid_to_variable_map.csv}} have mapped many chartevent d\_items from the two eras using column names, units of measure, data distributions, and contextual clues.
However, many 
% (how many?) 
d\_items have no known mapping between the two eras.
In practice, because the most important d\_items are believed to have been mapped, machine learning experiments usually ignore the unmapped features or learn distinct patterns in the two eras.
Because of changes in documentation practice related to the new EHR and temporal changes in patient makeup or medical practices, some shift in feature distributions between these eras is expected.

\subsection{Notation} \label{sec: notation_PCA}
Assume that we have multiple databases $\lbrace D^1, D^2, \ldots, D^I \rbrace$ to be schema-matched.
Indexing databases by $i$, each has $p^i$ features.
To simplify the notation, we will focus on the case where $I=2, i \in \lbrace A , B \rbrace$, avoiding the need to index many quantities that depend on the pair of databases.
When more than two databases need to be joined, one could directly apply pairwise methods in a round-robin fashion.
The $I=2$ case supports the most common application for medicine: a new database ($D^B$) joins an existing consortium ($D^A$).
%Without loss of generality, $D^B$ will have the larger number of features when the two are unequal.

We distinguish the database $D^i$ (which also contains meta-data) from the contained data $x^i$.
We will focus on the case that databases contain only numeric data (i.e., they are matrices in $\mathbb{R}^2$) although extensions to higher-order (tensor) databases and embedding complex objects are possible.
Denote an encoder for database $i$ as $f^i$ and a decoder $g^i$; that is for any example $j$, $x^i_{j} \approx \left(g^i \circ f^i\right) (x^i_{j}) $.
Denote by $L(,)$ a loss function appropriate to the data type at hand.
Because $f$ and $g$ operate on row-vectors, we will write them as functions of both tuples ($x^i_{j}$) and matrices ($x^i$) where unambiguous.
% Define the number of features in common between databases as $m$.
We require the output spaces of $f^A$ and $f^B$ to be identical in dimension, and we refer to $g^A \circ  f^B$ as the \textit{chimeric encoder} from $B$ to $A$. 
We also compute \textit{chimeric dependence} denoted by $r$ and defined as the dependence between every pair of features from $x^i$ and the output of chimeric encoder (when $x^i$ is given as an input). 

\subsection{Proposed algorithms} \label{sec: proposed_method}
This section presents the modules involved in the algorithm in order.
The complete algorithm is presented in Figure \ref{fig: Schema_full_2stage_Chimeric}.
The overall flow is as follows:
\begin{enumerate}
    \item Pre-process both databases to create a tabular representation without explicit missingness. Identify a small number of features known to correspond across the two databases using any of the other approaches to schema matching.
    \item \textit{KMF Module.} Compute an association vector between each unmapped feature and the set of mapped features. Use the cosine similarity of these vectors to create a similarity matrix between unmapped features in the two databases.
    \item \textit{Gale-Shapley Matching Module.} Transform the above similarity matrix to a discrete set of proposed mappings using the rankings of similarities as an affinity between two features.
    \item \textit{Chimeric autoencoder module.} Using the features mapped in the pre-processing and KMF modules, fit a pair of AE with a common latent space. Using the encoder from one dataset and decoder from the other, estimate paired samples of original data and data transformed to the format of the other database.
    \item \textit{Discover mappings in transformed data.} Create a similarity matrix between known-mapped features and transformed features using a bivariate association measure. Use the Gale-Shapley module to identify additional matches, surrogates, and transformed variables.
\end{enumerate}

%We propose two algorithms with varying level of complexity and their corresponding final usage in the schema matching and transformation pipeline. 
%The first one, referred to as the Known-Map Association Fingerprints (KMF) method, utilizes the correlation between the attributes. 
%The second one called Chimeric encoder based schema matching relies on identifying the common latent space for two data distributions, in addition to the correlation.
%Before describing the proposed algorithms in detail, we present data pre-processing requirements for the algorithms.

\subsubsection{Pre-processing}
In our real-data example, we use several pre-processing steps.
First, we assume that a subset of variables are pre-mapped in the two databases.
This can be based on the multiple sources of metadata discussed in the ``related work'' section above.
For continuous variables, although unit conversions can cause unexpected failures, nearly matching ranges, means, or variance are likely to be identical and this information can be included in the pre-matching stage.
Similarly, the numerous methods for matching text variables can be applied and that text can be transformed into continuous variables via neural embedding.
In practice, this would likely be followed by a manual verification stage.
Even if exploring all potential matches is impossible, checking some initial guesses with a content expert can likely be accomplished quickly.
We assume that $K$ variables are matched with certainty weights $w_k$ with $k \in [1, K]$. Also, for clarity in indexing, the known-mapped variables are moved to the front of each database in the same order.

We one-hot encode categorical variables in each dataset.
Our justification for this step is that categorical variables are semi-arbitrarily split and combined in real data sets, so forcing a $1:1$ match between databases may fail unexpectedly.
We unit-norm quantitative variables because, as mentioned above, we suspect that unit conversions occur commonly and because the mean and scale information are expected to have been used as part of the initialization of known-mapped columns. 
This step also simplifies the development of our neural network-based procedures, which can be adversely affected by inputs on very different scales.
Although not required, we greatly simplify the presentation by assuming that preliminary exploration (for example, using a single autoencoder) has been performed to detect and merge features that are identical concepts within a single dataset.
This assumption is necessary to define a ``correct'' mapping rather than an equivalence-class of features.
We also assume that missing data in each database has already been imputed; however, modifications to the proposal to allow missing data are straightforward.
In our example, we use imputation with predictive mean matching \cite{buuren_mice_2011} to fill in missing data.
For higher-dimensional objects like time series, an approach like Gaussian process adapters \cite{li_scalable_2016} can embed irregularly sampled data.

\subsubsection{Known-Map Association Fingerprints (KMF)}
As a fast alternative or initialization for the Chimeric encoder, we use a simple correlation-based approach.
We create a fingerprint of each unmapped column using the $K$-vector of correlation to the mapped columns.
We use the Pearson correlation coefficient because it is invariant to scale and centering, which are affected by unit of measure changes and assumed to have already been used in the verification of known mapped columns. 
The matrix of cosine similarity between fingerprints can then be estimated across databases and maximized by the matching algorithm below. 

\subsubsection{Mapping via Gale-Shapley}
After creating a similarity matrix between features, many methods have been used to infer a set of mappings, and there is no uniformly optimal way to do so \cite{shraga2020adnev}.
The Gale-Shapley algorithm is a widely-used algorithm for matching preferences of ``applicants'' and ``reviewers''  which has the advantage of depending only on the ranks of $r$ rather than absolute values \cite{iwama_survey_2008}. 
Before running Gale-Shapley, a-priori impossible matches (such as those with different types or extremely different marginal distributions) can be removed, as are variables with known matches.
Because of the extensive prior work on creating univariate summaries to use as a filter reviewed in Section \ref{sec: related_work}, we do not recapitulate that development here.
Gale-Shapley produces locally optimal matches (no pairwise trades are mutually beneficial) and is robust to misrepresentation of preferences by applicants, meaning that matches are never degraded by improving the optimization of $r$ or addition of irrelevant alternatives.
In the examples where the number of features is identical in $D^A$ and $D^B$, the ``stable marriage'' version of Gale-Shapley is used; otherwise, the ``hospital-resident'' version is used with a single acceptance slot per ``hospital''. 

After creating proposed mappings, we determine a threshold at which matches are so weak that they are likely false positives.
Because sample correlation has well-known statistical properties, we compute approximate $p$-values for each proposed mapping in a hold-out sample.
We use the Benjamini–Yekutieli procedure \cite{benjamini_yekutieli_2001} on the $p$-values to set an acceptance threshold which controls the false discovery rate under arbitrary correlation, as implemented by the Pingouin Python module \cite{vallat2018pingouin}. 
%For alternative measures of association, such as mutual information, permutation or subsampling based $p$-values should be used.

% and maximized by Gale-Shapley as above.
% Proposals above a threshold of similarity or a top fraction of proposals are then added to the ``known map'' set before running the chimeric encoder.

Since, KMF is computationally inexpensive, we identify the proposals (new matches) above a threshold of similarity or a top fraction of proposals and add them to the ``known map'' set before running the chimeric encoder in Section \ref{mainproposal}.
%(pictorial depiction available in Figure \ref{fig: Schema_full_2stage_Chimeric}).

\subsubsection{Chimeric translation} \label{mainproposal}
Even though KMF is simple and works efficiently for schema matching, it does not have the capability to identify surrogates for features with no match or to transform features. 
Therefore, we propose chimeric encoders that in addition to matching, can jointly transform and impute features between two databases.
The overall approach is illustrated in Figure \ref{fig: Schema_full_2stage_Chimeric}.

Define an autoencoder (AE) for each database using the notations in Section \ref{sec: notation_PCA}:
\begin{equation} \label{autoencoder}
    \hat{x}^i = g^i( f^i( x^i) ) , 
\end{equation}
and define a chimeric encoder:
\begin{equation} \label{chimeric}
    z^{i,i'} = g^{i'}( f^i( x^i) ) .
\end{equation}
In the 2-database case, we omit the redundant index and label $z^{i,i'}$ as $z^{i'}$.
With databases with very different numbers of features, a shared output dimension of the encoders can have unacceptably poor reconstruction accuracy, in which case an adapter function would need to be applied between $f^i$ and $g^{i'}$ in equation \eqref{chimeric}.
During training, the total loss is the reconstruction loss of each AE ($\sum_i L(\hat{x}^i , x^i)$) plus any regularization losses, plus a cross-reconstruction loss (encode, decode to other space) on the chimeric encoder which considers only the $K$ known-mapped features:
\begin{equation} \label{crossloss}
  \sum_{k=1}^K w_k L(z^{B}_{k} , x^B_{k}),
\end{equation}
plus the symmetric expression in $A$.
We also add a cycle-consistency loss (encode, decode to the other space, re-encode, decode to original space):
\begin{equation} \label{cycleloss}
  L( g^B( f^A( g^{A}( f^B( x^B) ) ) ), x^B ),
\end{equation}
plus the symmetric expression reconstructing $X^A$.
The paired samples $(x^i,z^{i'})$ are the key inputs to the next step of the algorithm.

The rationale for the loss in equation \eqref{crossloss} is that encoded space elements which are important to reconstruct the $K$ mapped features will also tend to be important in reconstructing the unmapped variables, similar to the latent components in principal components analysis.
Requiring that the two encoded spaces have similar relationships to the known-mapped columns will hopefully encourage that they have similar relationships to the unmapped columns.
The cycle consistency loss \eqref{cycleloss} also encourages the encoded spaces to have similar or identical semantics because they are used to cross-decode across databases.
Unlike the cross-reconstruction loss \eqref{crossloss}, the cycle consistency loss includes structure within the unmapped columns.

%The cross-reconstruction loss considers only the reconstruction of pre-mapped variables and assumes (wlog) that these have been placed in order at the first columns of the data matrices.
%The above formulation assumes that a single pre-match has been created for each variable, but with messier indexing and notation multiple candidates can be accommodated.
We consider regularization losses including weight decay and orthogonalization of the latent space \cite{chang_scalable_2018}, but we do not present here due to space constraints.
The encoders are given substantial noise injection (Bernoulli or Gaussian dropout) and a narrow latent dimension to encourage the latent space to collapse the data to a shared representation (i.e. requiring information from each component to be used for reconstructing multiple columns).

\subsubsection{Chimeric mapping}
Having created paired samples $(x^i, z^{i'})$, we then have the task of identifying feature correspondence. 
% We propose to create a similarity matrix using a measure of dependence for each pair $(k,l)$ of columns in $(x^i, z^{i'})$ (which we refer to as \textit{chimeric dependence}), $r_{kl}$. 
We propose to create a similarity matrix using chimeric dependence, $r_{kl}$, for each pair $(k,l)$ of columns in $(x^i, z^{i'})$. 
As explained earlier, we use Pearson correlation 
% is invariant to scale and center, which are the most common recording discrepancies and distributional shifts between databases, and we therefore use it 
as $r$ for our experiments to detect features which are mapped without any transformation.
Alternatives such as mutual information are considered in the discussion section and are more appropriate to detect features which are transformed.
We then use the same Gale-Shapley algorithm on the chimeric dependence as we described for KMF.
% The Gale-Shapley algorithm is a widely-used algorithm for matching preferences of ``applicants'' and ``reviewers''  which has the advantage of depending only on the ranks of $r$ rather than absolute values \cite{iwama_survey_2008}. 
% Before running Gale-Shapley, a-priori impossible matches (such as those with different types or extremely different marginal distributions) can be removed, as are variables with known matches.
% Because of the extensive prior work on creating univariate summaries to use as a filter reviewed in Section \ref{sec: related_work}, we do not recapitulate that development here.
% Gale-Shapley produces locally optimal matches (no pairwise trades are mutually beneficial) and is robust to mis-representation of preferences by applicants, meaning that matches are never degraded by improving the optimization of $r$ or addition of irrelevant alternatives.
% In the examples where the number of features is identical in $D^A$ and $D^B$, the ``stable marriage'' version of Gale-Shapley is used; otherwise the ``hospital-resident'' version is used with a single acceptance slot per ``hospital''. 
A step by step procedure including both Chimeric encoding and Gale Shapley matching is given in \ref{alg: Full_mapping}.
%In this setting too, we performed the Benjamini–Yekutieli procedure to finalize the matches. 
% After creating proposed mappings, we determine a threshold at which matches are so weak that they are likely false positives.
% Because sample correlation has well-known statistical properties, we compute approximate $p$-values in the hold-out sample for each proposed mapping.
% We use the Pingouin \cite{vallat2018pingouin} module from Python to perform the Benjamini–Yekutieli procedure \cite{benjamini_yekutieli_2001} on the $p$-values to set an acceptance threshold which controls the false discovery rate under arbitrary correlation. 
% For alternative measures of association, such as mutual information, permutation or subsampling based $p$-values should be used.
We considered a threshold in the absolute similarity in addition to the Benjamini–Yekutieli based stopping; however, we found that the high degree of compression in our autoencoder networks tended to degrade these correlations from the ideal and that a single threshold could not easily be selected.
Gale-Shapley is an asymmetric matching process; a given application may provide a rationale to favor one database over the other (such as $D^A$ being much larger); alternatively, in a symmetric problem this procedure can be repeated with $(x^i, z^{i'})$ from each $i$, and either the similarity matrices are combined prior to mapping or the mappings are compared for consistency.
The optimal way to combine similarity matrices is itself a topic of active research \cite{shraga2020adnev}. 
In our examples, we arbitrarily use the $(x^A, z^{B})$ mapping for simplicity.

\begin{algorithm}[!htbp]
		\renewcommand{\thealgorithm}{Algorithm AE\_GS\_matching}
		\floatname{algorithm}{}
		\caption{Chimeric schema matching with $k$ pre-mapped features}
		\label{alg: Full_mapping}
		\begin{algorithmic}[1]
		\Statex Mini-batch inner loop within epochs not shown for readability. Also omitted any regularization losses such as weight decay and orthogonalization.
			\Statex \hspace{-0.44cm}\textbf{Input:} Training data: $x^A$ (feature set $\mathcal{F}_A$ ), $x^B$ (feature set $\mathcal{F}_B$), List of mapped features $\mathcal{F}_m$ of size $k$ 
			\Statex \hspace{-0.44cm}\textbf{Require:} Learning rate $\alpha$, max epochs $n_e$,  Weight for cross loss $w_c$, Weight for cycle consistency loss $w_{cy}$, initial parameters $\theta_A$, $\theta_B$, Matching algorithm \textbf{GSM}.
			\Statex \hspace{-0.44cm}\textbf{Output:} Mappings between $\mathcal{F}_A$ and $\mathcal{F}_B$, i.e., $M_{x^A}$, $M_{x^B}$.
			%\Statex \hspace{-0.44cm}\textbf{Define:}\textit{Direct Rec loss}: $aeA_d$, $aeB_d$,  \textit{Cross loss}: $aeA_{c}$, $aeB_{c}$, \textit{Cyc. cons. loss}: $aeA_{cy}$, $aeB_{cy}$ \textit{Orthg loss}: $orthA$, $orthB$

			\State Let $(f^A, g^A)$ and $(f^B,g^B)$ be the encoder and decoder of $x^A$ and $x^B$ 
% 			\For{$t = 0,\cdots, n_{e}$} \COMMENT{[Initialize AE]}
% 			\State  $AE_{A}(\theta_A) = \mathbf{L}(x^A,g^A(f^A(x^A)))$, \quad $AE_{B}(\theta_B) = \mathbf{L}(x^B,g^B(f^B(x^B)))$
% 			\State loss $= (AE_A + AE_B) $
% 			\State  $\theta_A = \theta_A - \alpha*\nabla(\textrm{loss}, \theta_A)$ \COMMENT{[Or Adam, other updates]}
% 			\State  $\theta_B = \theta_B - \alpha*\nabla(\textrm{loss}, \theta_B)$
% 			\EndFor 
			\For{$t = 0,\cdots, n_{e}$}
			%\For{$mb\_x^A, mb\_x^B$ \textbf{in} $x^A,x^B$}
			\State  $AE_{A}(\theta_A) = \mathbf{L}(x^A,g^A(f^A(x^A)))$, \quad $AE_{B}(\theta_B) = \mathbf{L}(x^B,g^B(f^B(x^B)))$
			\State $CE_{A}(\theta_A,\theta_B) = \mathbf{L}(x^A[,1:k],g^B(f^A(x^A)))[,1:k]$, \quad $CE_{B}(\theta_A,\theta_B) = \mathbf{L}(x^B[,1:k],g^A(f^B(x^B))[,1:k])$
			\State $CY_{A}(\theta_A,\theta_B) = \mathbf{L}(x^A,g^A(f^B(g^B(f^A(x^A)))))$, \quad $CY_{B}(\theta_A,\theta_B) = \mathbf{L}(x^B,g^B(f^A(g^A(f^B(x^B)))))$
%            \State  $orthA = \Vert f^A(x^A)^T f^A(x^A) - \mathbb{I} \Vert$, \quad $orthB = \Vert f^B(x^B)^T f^B(x^B) - \mathbb{I}\ \Vert$
			\State loss $= (AE_A + AE_B) + w_c(CE_A + CE_B) +w_{cy}(CY_A + CY_B) $
			\State  $\theta_A = \theta_A - \alpha*\nabla(\textrm{loss}, \theta_A)$ 
% 			\COMMENT{[Or Adam, other updates]}
			\State  $\theta_B = \theta_B - \alpha*\nabla(\textrm{loss}, \theta_B)$
			%\EndFor
			\EndFor 
			\State $CC_{x^A}$ = Cor$(x^A,g^B(f^A(x^A)))$, $CC_{x^B}$ = Cor$(x^B,g^A(f^B(x^B)))$
			\State $M_{x^A} = \mathbf{GSM}(CC_{x^A})$, $M_{x^B} = \mathbf{GSM}(CC_{x^B})$
			
		\end{algorithmic}
\end{algorithm}

\paragraph{Hyperparameter tuning}
We recommend a leave-one-out approach to select the architecture of encoders and decoders, regularization strength, learning rate, and other hyperparameters in \ref{alg: Full_mapping}.
That is, after verifying the $K$ known mapped features, mark only $K-1$ as known, run \ref{alg: Full_mapping}, and evaluate the mapping accuracy on the held-out feature in $K$ runs.
As in any hyperparameter tuning, prior knowledge from related tasks and iterative evaluation of the results in easily verified cases are likely to narrow the search space and reduce computational costs.

\paragraph{Federated learning}
We note that the above chimeric schema matching algorithm does not require any party to have direct access to more than one database.
Each database can be hosted on a distinct server and pass current parameters of $f$ and $g$ for parameter updates, gradient calculations, or partial loss function calculation using the other databases.

\paragraph{\textbf{End-to-end schema matching procedure}}
The final schema matching procedure is a combination of the above two methods where we use KMF as an initialization step to increase the number of pre-mapped features and then run the chimeric encoder to obtain the final mapping and transformations. A schematic diagram of the above combination is depicted in Figure \ref{fig: Schema_full_2stage_Chimeric}.

% \begin{figure*}[h!]
%     % \centering
%     % \begin{subfigure}[b]{0.478\textwidth}
%         \centering
%         \includegraphics[width=1\textwidth]{Figures_final/KMF-ChimericAlgo_Schematic-cropped.pdf}
%         \caption{An illustration of the KMF initialized chimeric schema matching algorithm with module-wise description at the start of Section \ref{sec: proposed_method}. It also depicts how the input datasets could be differently sized with few mapped features (first three blue coloured columns with common feature names) and majority unmapped (pink and orange coloured columns with ambiguous feature names).}
%         % {}    
%         \label{fig: Schema_full_2stage_Chimeric}
% \end{figure*}

We envision that the above process would be iterative.
Having created candidate matches, these could be verified using alternative data sources.
With these new ``known'' variables the algorithm can be restarted, bootstrapping to identify more matches. 
Having assigned maps to most features from $D^A$ to $D^B$, the translated value would be an element-wise combination of the mapped variable (where defined) and the chimerically encoded value otherwise.

We also evaluated a simpler supervised learning approach to generating paired $(x^i, z^{i'})$ samples. 
First, fit a classifier or regression function for the unmapped features in each dataset as a function of the mapped features only, $x^{i'} \sim h^{i'}(x^{i'}_{1:k})$, second use the fitted function from one dataset on samples from the other dataset $h^{i'}(x^i_{1:k})$ to generate $z^{i'}$ by parametric simulation or predictive mean matching to a sample in $x^{i'}$.
This approach does not take advantage of any structure within the unmapped features, but is much easier to fit.
However, we did not find any performance advantages over the full chimeric encoder, and its results are not reported.

\subsection{Experimental details}
%In this section, we discuss the data generating mechanism for synthetic datasets, real dataset details, organization of experiments, and finally the metrics used to assess the performance of the algorithms.
\subsubsection{Synthetic Data generation}
We consider three synthetic data-generating mechanisms: multivariate Gaussian,
%where the mean vector for each dimension was randomly sampled from  $[a,b]$, 
 2-cluster multivariate Gaussian, and a binarized version of the 2-cluster multivariate Gaussian. % where each $n$-dim Gaussian has mean vectors sampled from  $[a,b]$ separately. 
The mixture case is intended to mimic data derived from a case-control study.
Details of the simulation can be found in the Appendix \ref{synth_details} and a code repository in \ref{codeappendix}.
Samples for each database are drawn independently from these distributions.
We consider simulations where the two column spaces are identical, one is a subset of the other, and where they only partially overlap by randomly dropping columns from one or both datasets respectively.
Known mapped columns are assigned randomly from the set of shared columns.
The remainder (unmapped columns) have their order randomly permuted.
In some experiments, we transform a randomly selected column by squaring it, making it zero correlation with the raw data.
Finally, to check the null behavior of our models, we generate data according to independent Gaussians, from which no information for matching should be available. The results for this experiment are provided in Appendix  \ref{app: null_case}, but in no settings did the algorithm produce less than the maximal number of mismatches.

\subsubsection{Real data description}

We used the MIMIC-III v1.3 dataset \cite{johnson2016mimic} to demonstrate our algorithm for mapping features between two eras, Carevue (CV) and Metavision (MV).
%specifically, continuous valued features corresponding to labs and chartevents. 
We included patients who were 18 years and older at the time of hospital admission and survived beyond the first 24 hours of their first ICU stay. 
Where more than 1 ICU stay occurred, we included only the first.
We treated the laboratory table (labevents) as the known-mapped columns, and used the nurse-charted data table (chartevents) as the unmapped columns.
Features were identified as belonging to the CV or MV era based on the value of their d\_items index. A small number of features appeared to substantially overlap the two eras and were removed from the unmapped set.
Some labevents data is directly copied into chartevents, and these features were removed from chartevents.
Because both tables are time series, we selected the last observation in the first 24 hours for each patient, excluding any observations tagged as errors.
%We did not include the observations that were tagged as error in MV era data. 
%We removed the chartevent features that originally belonged to MV era but existed in CV era data too. 
We excluded features with no data in more than $90\%$ of patients in labevents and $80\%$ in chartevents. 
After these restrictions, the CV dataset had $26130$ patients, and the MV dataset had $21125$ patients. Both datasets had $71$ chartevents features and $58$ labevents features.
Of the included chartevents features, 49 have been mapped by prior work, and these mappings are treated as a ``gold standard'' to evaluate the algorithm.
The remaining 22 features are included in the procedure, but the results are not included in the evaluation.
We treat mapping a feature with a gold-standard pairing to another feature as an error, but mapping between two features with no gold standard are ignored.
In experiments varying the number of mapped features, features are selected at random from the labevents table.
Labevents features not included as mapped are included in the evaluation set of unmapped features.

We used the MIMIC dataset to validate our proposal for hyperparameter selection for the chimeric encoder method.
Because the number of mapped features ($k$) in MIMIC dataset was large, we chose to randomly split the sample of lab features into half used for training (marked as known in the procedure) and hold-out (used to evaluate F1 score). This cross-validation split was chosen instead of leave-one-out because with increase in $k$, using leave-one out strategy for hyperparameter tuning would become computationally expensive and the estimates would have higher variance.
% Because the number of mapped features ($k$) in MIMIC dataset was large, we chose to randomly split the sample of lab features into half used for training (marked as known in the procedure) and hold-out (used to evaluate F1 score). 
% % This cross-validation split was chosen instead of leave-one-out because the performance appeared to have saturated.
% This cross-validation split was chosen instead of leave-one-out because when $k$ is large, it will be an easy task for any $k-1$ sized mapped feature subset to match the one remaining unmapped feature.
% We noticed the implication of this behaviour in the form of performance saturation.
10 such cross-validation folds were evaluated for each hyperparameter set (grid search) and the true accuracy on the chartevents table compared between the cross-validation selected hyperparameters and the global optimizer.

For the ACTFAST dataset, we included all  binary or categorical features from the preoperative evaluation. The dataset had $97068$ patients with $31$ binary features and $9$ categorical features. Feature names are available in Table \ref{tab: real_Data_feature_names} in Appendix \ref{app: real_Data_details}. Meta data and more details about the features can be found in  \cite{fritz_deep-learning_2019}. 
We always select the mapped features from the set of categorical features, since it seems more realistic that these text values would be pre-mapped by meta data. 
All categorical features are one hot encoded before analysis and therefore the total number of features is $81$ and the number of binary mapped features can vary with the selection of categorical features. 
Most of these binary variables have a low frequency of positive values, and no missing data was present in these features. 
Two experimental databases were created by randomly partitioning the patients and randomly dropping and permuting unmapped columns, simulating two ``eras'' in an identical manner to the synthetic data experiments.
%Unfortunately, the underlying record only recorded ``yes'' for most binary features and did not distinguish between ``did not answer'' and ``no.''
%``Did not respond'' was added as a level to categorical features.
% The experiment setup is the same as the synthetic dataset described in Section \ref{subsubsec: Exp_org_syn}. 

\subsubsection{Implementation and Baseline Methods}
Details of the neural network model architecture for algorithm \ref{alg: Full_mapping} and hyperparameter search for each experiment are contained in Appendix \ref{subsubsec: Model_arc_syn}.
We use PyTorch to implement all the neural network steps, and code is made available in Appendix \ref{codeappendix}. 

We compare Algorithm \ref{alg: Full_mapping} and the simplified version stopping after KMF initialization to two baselines in instance-based schema matching.
First, the mutual-information-similarity method of Kang and colleagues \cite{kang_schema_2008} is labeled ``Kang'' in plots and tables.
Unfortunately, there was no publicly available implementation of the Kang methods and we received no reply from the corresponding author.
We therefore include a re-implementation of their method in our code.
Second, RadialGAN \cite{yoon_radialgan_2018} transforms samples between database formats. Although it does not explicitly match features, we can feed its output into the same similarity-matrix and mapping steps in \ref{alg: Full_mapping}.
The authors of RadialGAN also declined to share code or data, and we have done our best to re-implement their approach.

\subsubsection{Metrics}
Our primary evaluation metric is F1 score computed on the accepted mappings between previously unmapped features. 
%This was chosen because, where a map exists, identifying it has a dramatic effect on overall performance.
%Matching accuracy also reflects how well the algorithm maintains alignment in the latent spaces.
The average F1 scores from the proposed methods and the baseline are compared with the Wilcoxon rank sum test.
In experiments where columns are intentionally dropped or transformed, we quantify the results where no match exists (imputation) or a transformation occurs with the correlation between the reconstructed value and the true masked value.
Because we generally think of the method as being applied to matching a large local pool of candidate variables to a smaller number in an existing consortium, we report matching with the larger number of columns as ``reviewers'' and the smaller number as ``applicants'', optimizing the ``preferences'' of the consortium more.   
Alternatively, the two sets of proposals could be combined or compared.

We considered augmenting classifier training data as an evaluation metric similar to Yoon and colleagues \cite{yoon_radialgan_2018}; however, we found this to require repeated tuning on the sample sizes, classifier complexity, and problem difficulty to prevent the classifier from effectively saturating its learning and therefore not improving much from additional data. If the primary dataset is small, the results on combined data quickly converge to simply using the external data. For example, logistic regression weights (coefficients) converge to population values with error $\propto n^{-\frac{1}{2}}$, and therefore at realistic sample sizes only modest changes occur with additional data.

\section{Results} \label{sec: experiment_results}
%In this section, we demonstrate the performance of the proposed Chimeric Autoencoder and the associated schema matching solution on few synthetic datasets and a real world ACTFAST dataset. In case of synthetic datasets, we start with the effect of varying factors like sample size and difference between the number of features between two databases on the performance of the chimeric approach. For the basic performance demonstration, we compare our proposed algorithms viz., KMF and KMF initialized chimeric encoder  with RadialGAN and MI based schema matching method by Kang and colleagues \cite{kang_schema_2008} (referred to as Kang method in this section) on both synthetic and real datasets. 

% Flow charts for experimental steps in both permutation and incomplete feature subset case are available in Appendix \ref{app: flow_chart_exp_procedure}.

\subsection{Synthetic data results}
Table \ref{tab:significance_test_results} presents the performance evaluations with statistical tests over all experiments.

\begin{figure}[h!] 
    \centering
    \begin{subfigure}[b]{0.478\textwidth}
        \centering
        \includegraphics[width=1.0\textwidth]{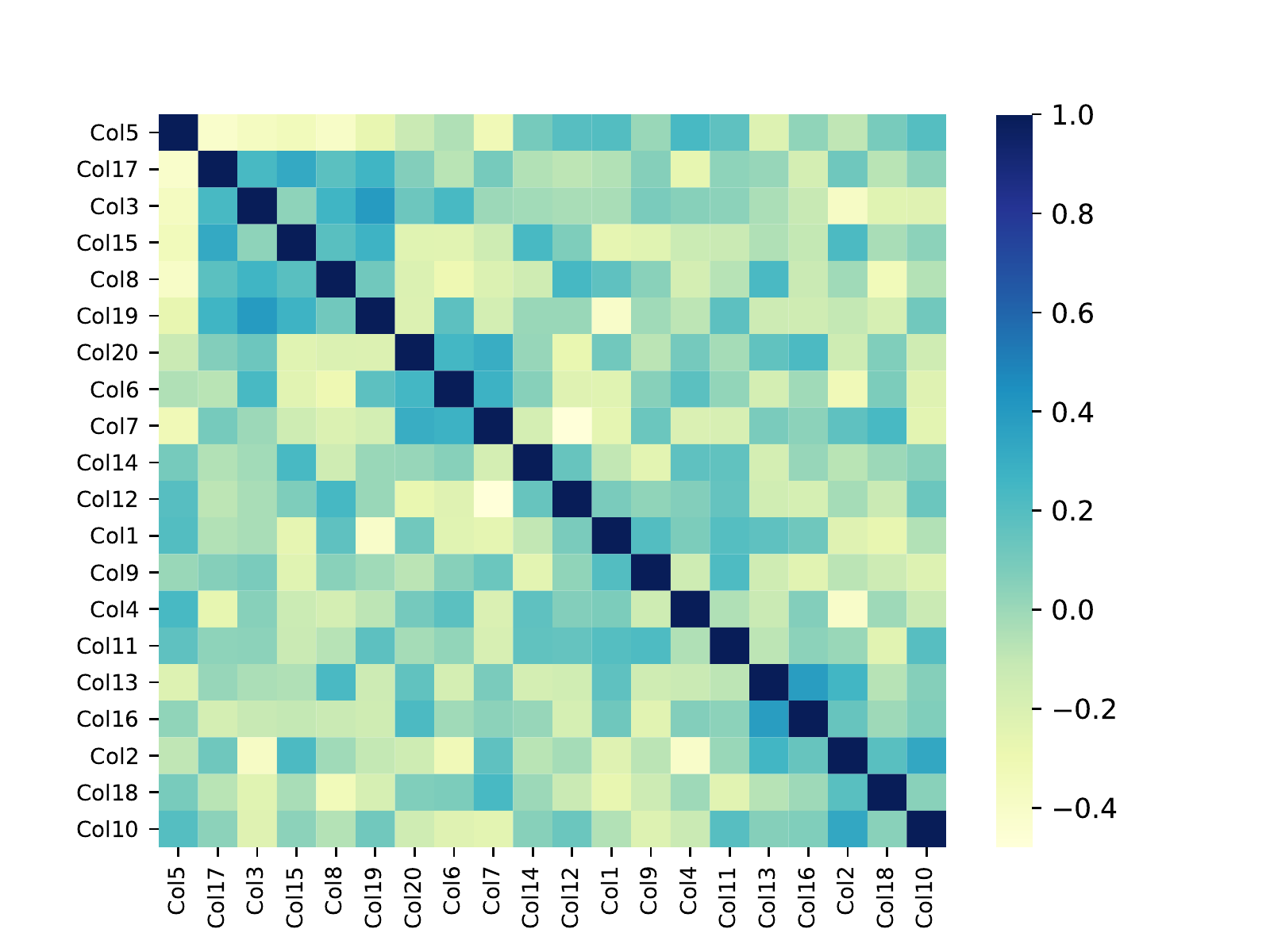}
        \caption{\footnotesize{}}
        % {}    
        \label{fig: SD2_correlation_clustered}
    \end{subfigure}
    \hfill
    \begin{subfigure}[b]{0.478\textwidth}  
        \centering 
        \includegraphics[width=1.0\textwidth]{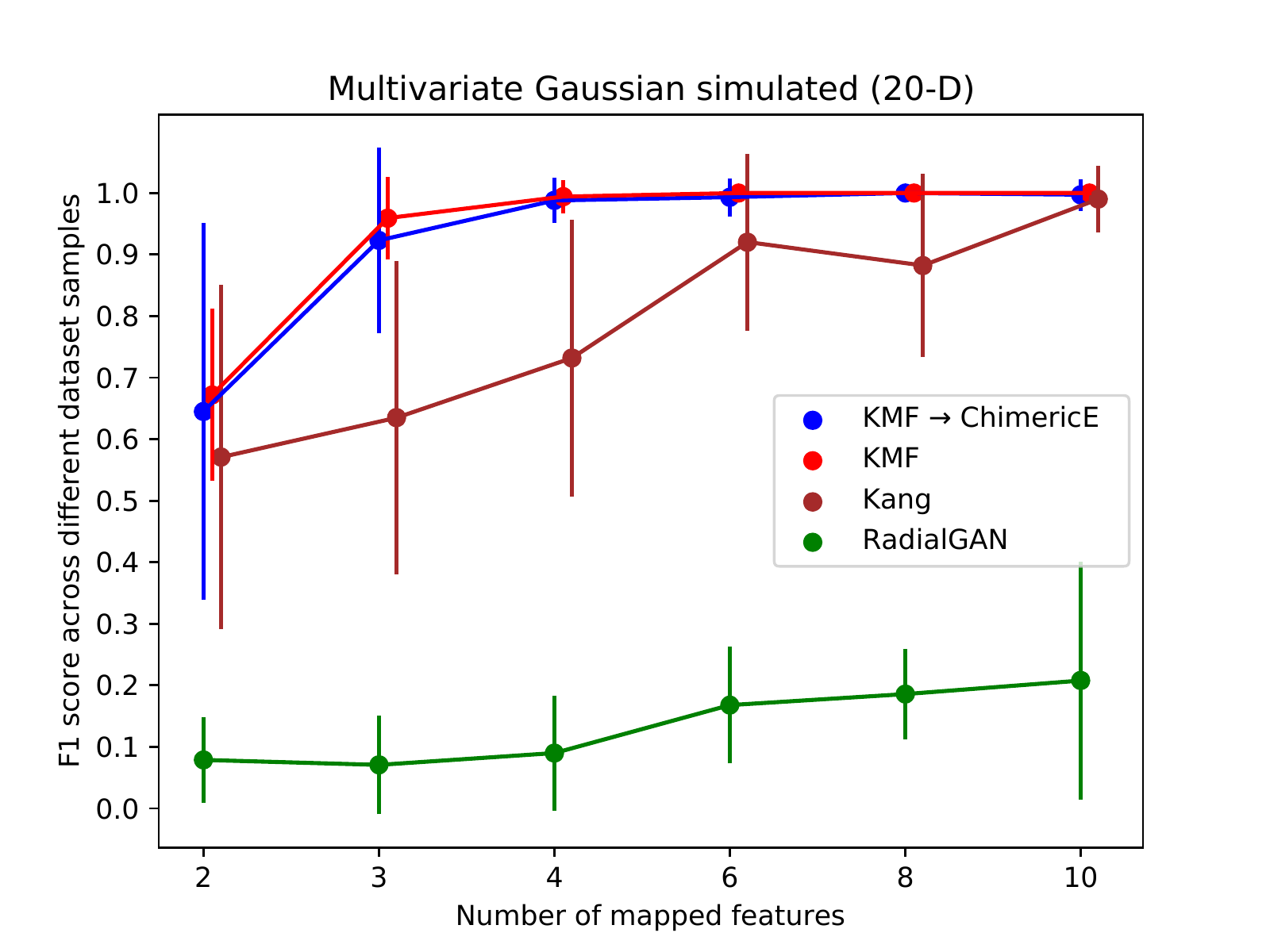}
        \caption{\footnotesize{}}
        % {}       
        \label{fig: Syn2_comp_Radial_GAN_Chimeric_vs_mappedfeatures}
    \end{subfigure}
    \vskip\baselineskip
        \begin{subfigure}[b]{0.478\textwidth}
        \centering
        \includegraphics[width=0.98\textwidth]{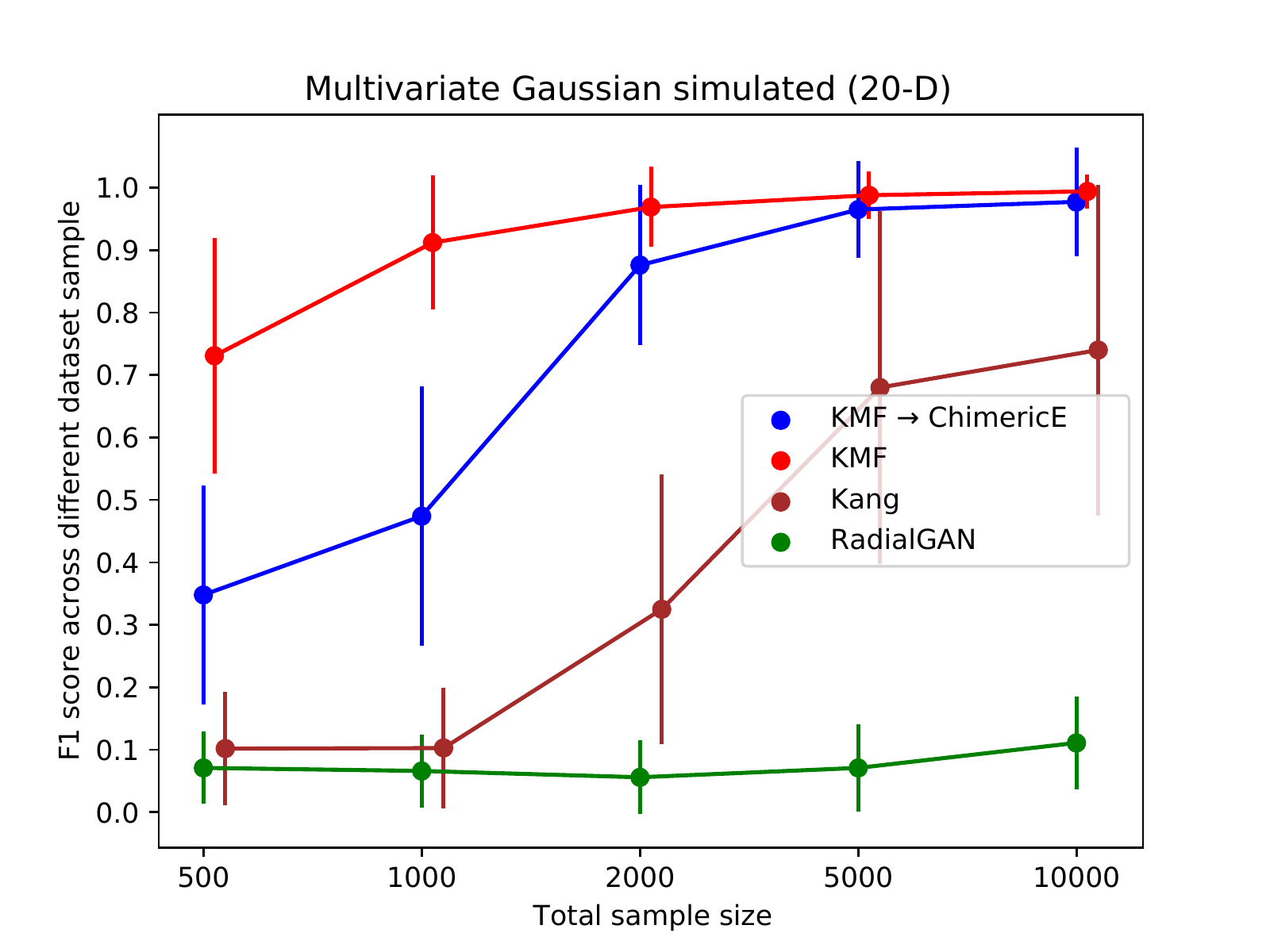}
        \caption{\footnotesize{}}
        % {}    
        \label{fig: Syn2_comp_Radial_GAN_Chimeric_vs_sample_size}
    \end{subfigure}
    \hfill
    \begin{subfigure}[b]{0.478\textwidth}  
        \centering 
        \includegraphics[width=0.98\textwidth]{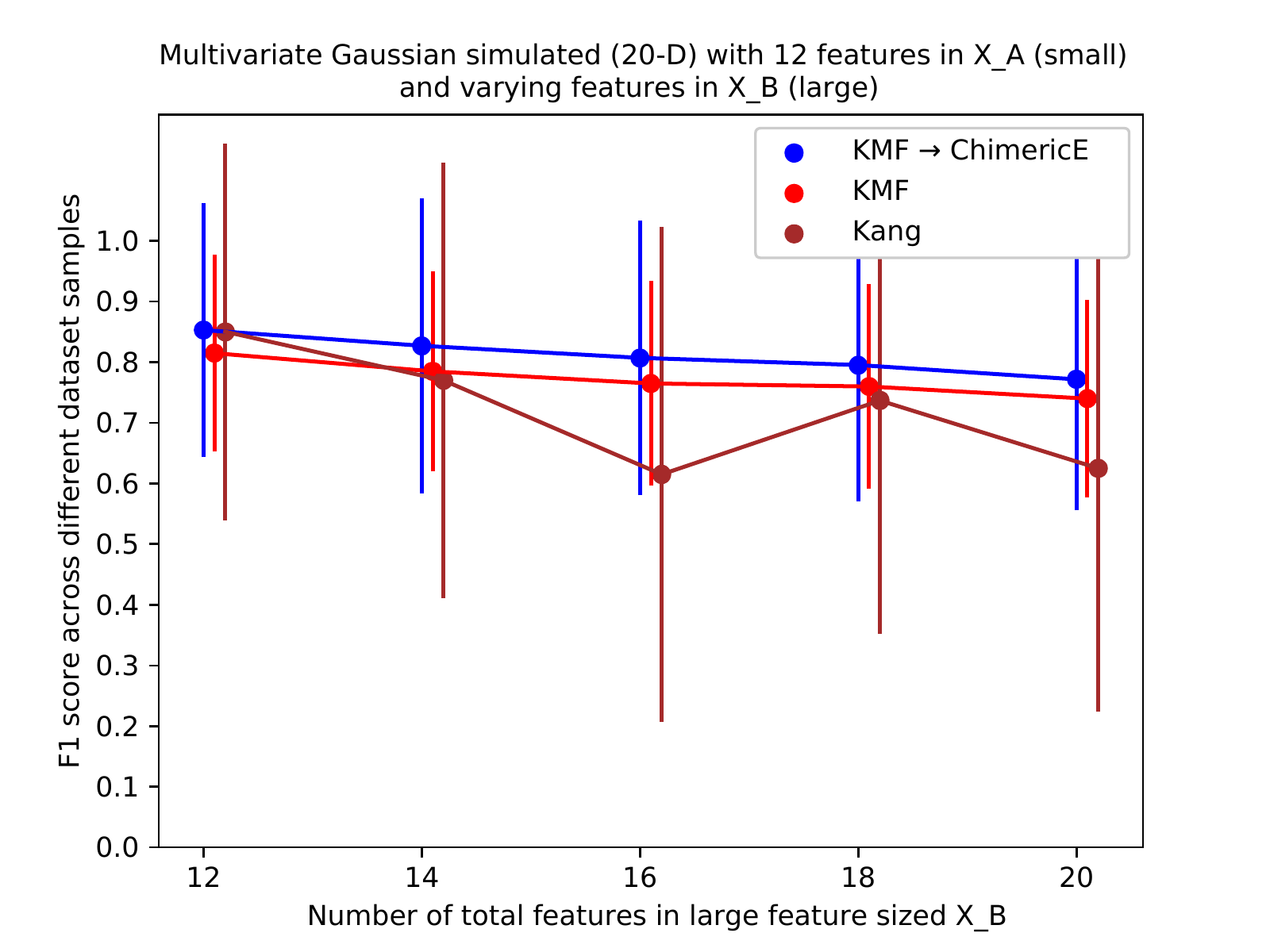}
        \caption{\footnotesize{}}
        % {}       
        \label{fig: Syn2_comp_KMF_Chimeric_vs_mapped_features_Onto_case}
    \end{subfigure}
    
    \caption{Comparison between KMF (red), KMF initialized chimeric encoder (blue), Kang (brown) and RadialGAN (green) on \textbf{Multivariate Gaussian simulated (20-D)} dataset (a) Correlation structure among features obtained after spectral clustering with 5 clusters (b) Performance variation as the number of mapped features increase and as expected the average F1 score increases with more prior information (c) Performance variation with increase in the total sample size with 4 pre-mapped features (d) Performance variation as the feature size increase in $x^B$ (onto map case) with 2 pre-mapped features.}
    \label{iid_fig}
\end{figure}

Figure \ref{iid_fig} shows the results in the simplest synthetic dataset \textbf{Multivariate Gaussian simulated (20-D)}. 
%Because RadialGAN and Kang do not have any mechanism of using prior knowledge of mapped features, in each run we feed all algorithms with exactly same data including already aligned mapped features at the start of the table. 
%The Kang method is initialized with the known elements of the permutation correct and these are set to never change.
Figure \ref{fig: Syn2_comp_Radial_GAN_Chimeric_vs_mappedfeatures} shows the performance as the number of mapped features varies.  
First, the RadialGAN method performs poorly under all conditions, and we will not repeat this finding with each scenario.
Second, the KMF and chimeric encoder quickly saturate the task, and Kang also performs well once more than 25\% of the features are mapped. 
The high standard deviation in Figure \ref{fig: Syn2_comp_Radial_GAN_Chimeric_vs_mappedfeatures}  is due to the selection of different mapped features across different trials. 
As can be seen in Figure \ref{fig: SD2_correlation_clustered}, there are only a few features with strong correlation to the rest of the features, and if the mapped feature set includes them then the F1 score for that trial is high. 
%This validates our hypothesis that chimeric encoders (and KMF) are exploiting the correlation between the features and accordingly generates a shared and informative latent representation (fingerprints). 

In  Figure \ref{fig: Syn2_comp_Radial_GAN_Chimeric_vs_sample_size}, we vary the sample size of the total dataset with the number of mapped features fixed at 4. 
For KMF and chimeric encoder methods, performance stabilizes around $5000$ samples, suggesting the variation that we observe in other experiments (using 10000 samples) is the large sample behavior. 
The chimeric encoder performs worse than KMF at small sample sizes, which is expected given the potential for its autoencoders to overfit.
The sample-size dependence of Kang is less easy to explain.
In Figure \ref{fig: Syn2_comp_KMF_Chimeric_vs_mapped_features_Onto_case}, we evaluate onto mapping (one feature set is a subset of the other) varying the number of unshared features. 
The chimeric encoder slightly outperformed others, but differences were small.  
Interestingly, increasing the number of candidate features had minimal effect on the F1 score.
 
% Finally, to test the time it takes for the shared latent representation to stabilize, we evaluate the model performance across different training epochs with the number of mapped features being fixed at $4$. Chimeric AE stabilizes very fast at about 20 epochs in comparison to RadialGAN which even after 1000 epochs is not able to get good cross reconstruction for getting the matchings correct.

\begin{figure}
    \centering

    \begin{subfigure}[b]{0.475\textwidth}
        \centering
        \includegraphics[width=0.8\textwidth]{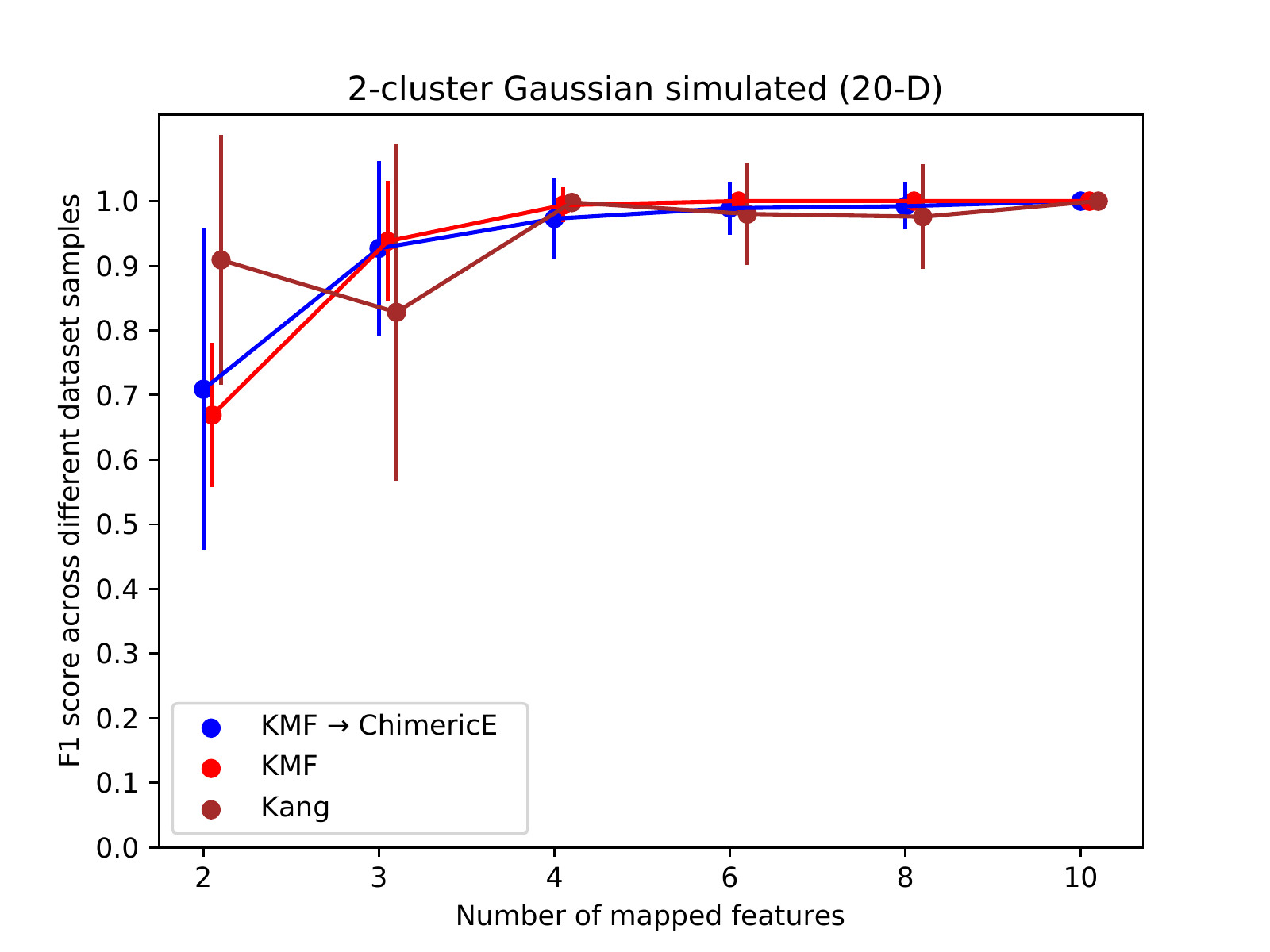}
        \caption{\footnotesize{}}
        % \caption{\footnotesize{\textbf{SD1}: Comparison wrt number of mapped features}}
        \label{fig: Syn1_Chimeric_vs_mappedfeatures}
    \end{subfigure}
    \hfill
            \begin{subfigure}[b]{0.475\textwidth}
        \centering
        \includegraphics[width=0.8\textwidth]{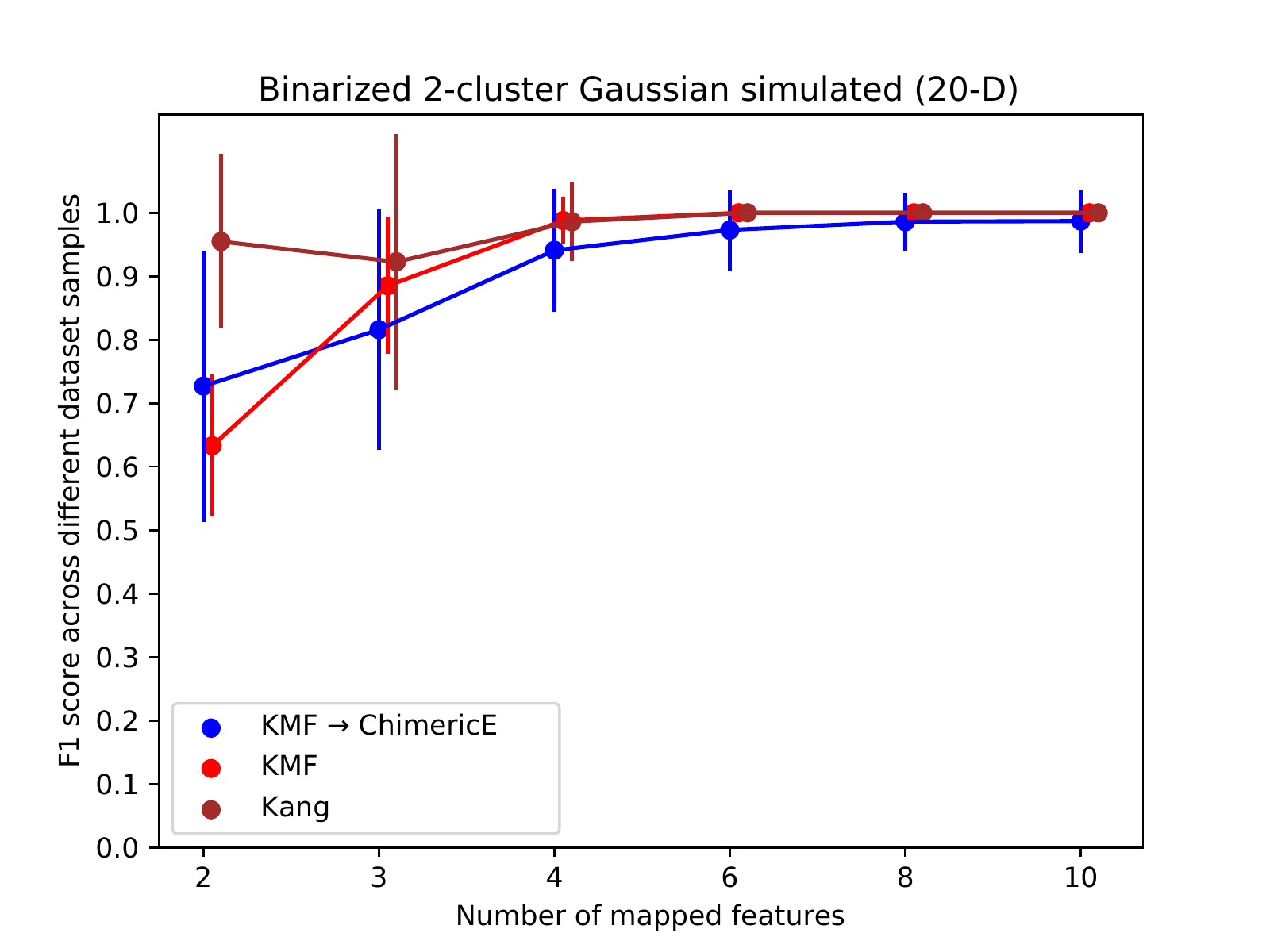}
        \caption{\footnotesize{}}
        % \caption{\footnotesize{\textbf{SD1}:Comparison wrt number of mapped features when the whole data was binarized after preprocessing}}
        \label{fig: Syn1_Chimeric_vs_mappedfeatures_binarized}
    \end{subfigure}
        \vskip\baselineskip
        
        \begin{subfigure}[b]{0.475\textwidth}
        \centering
        \includegraphics[width=0.8\textwidth]{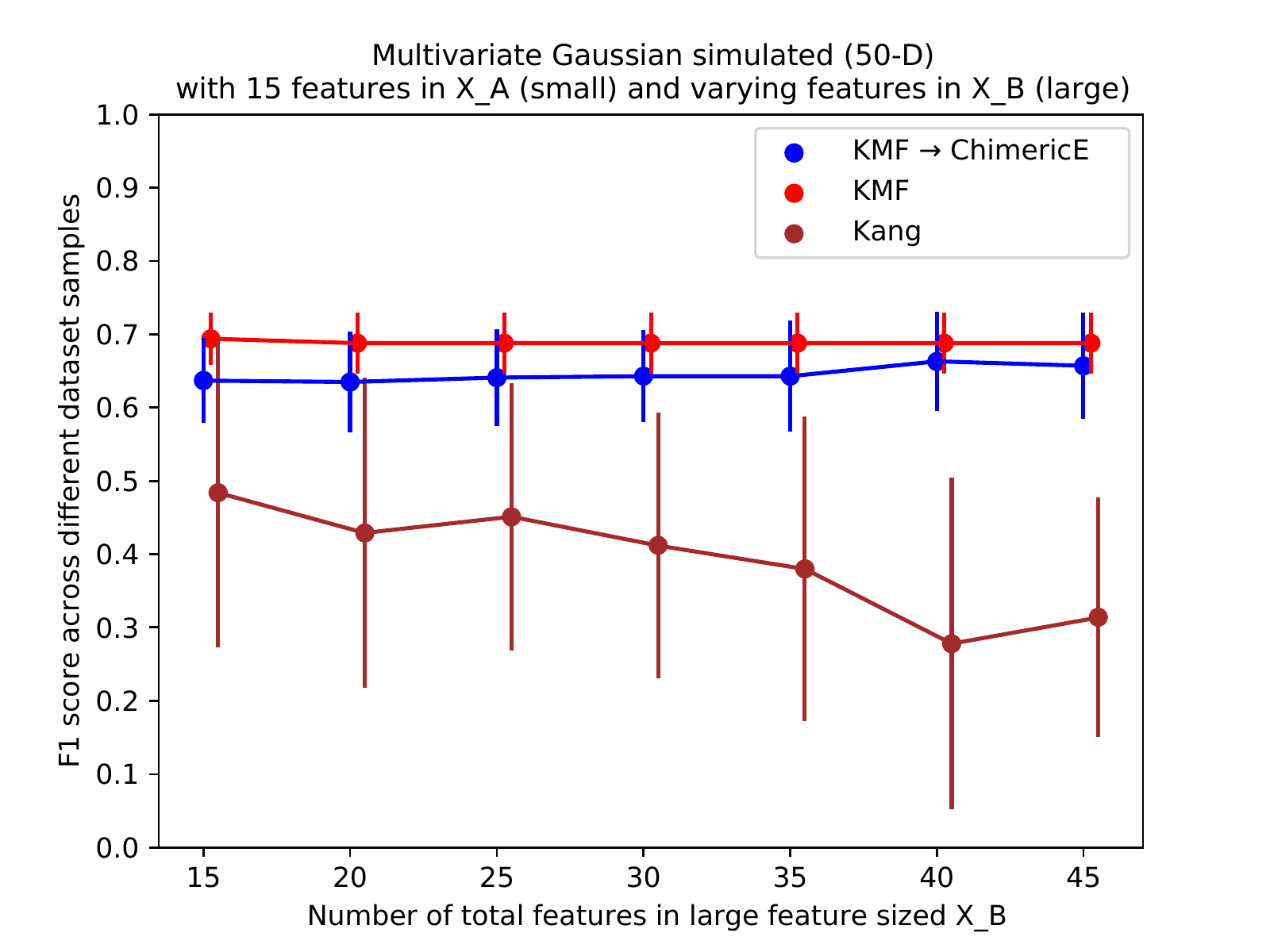}
        \caption{\footnotesize{}}
        \label{fig: Syn5_Chimeric_vs_increasing_number_of_features_inX2_PM}
    \end{subfigure}
    \hfill
        \begin{subfigure}[b]{0.475\textwidth}
        \centering
        \includegraphics[width=0.8\textwidth]{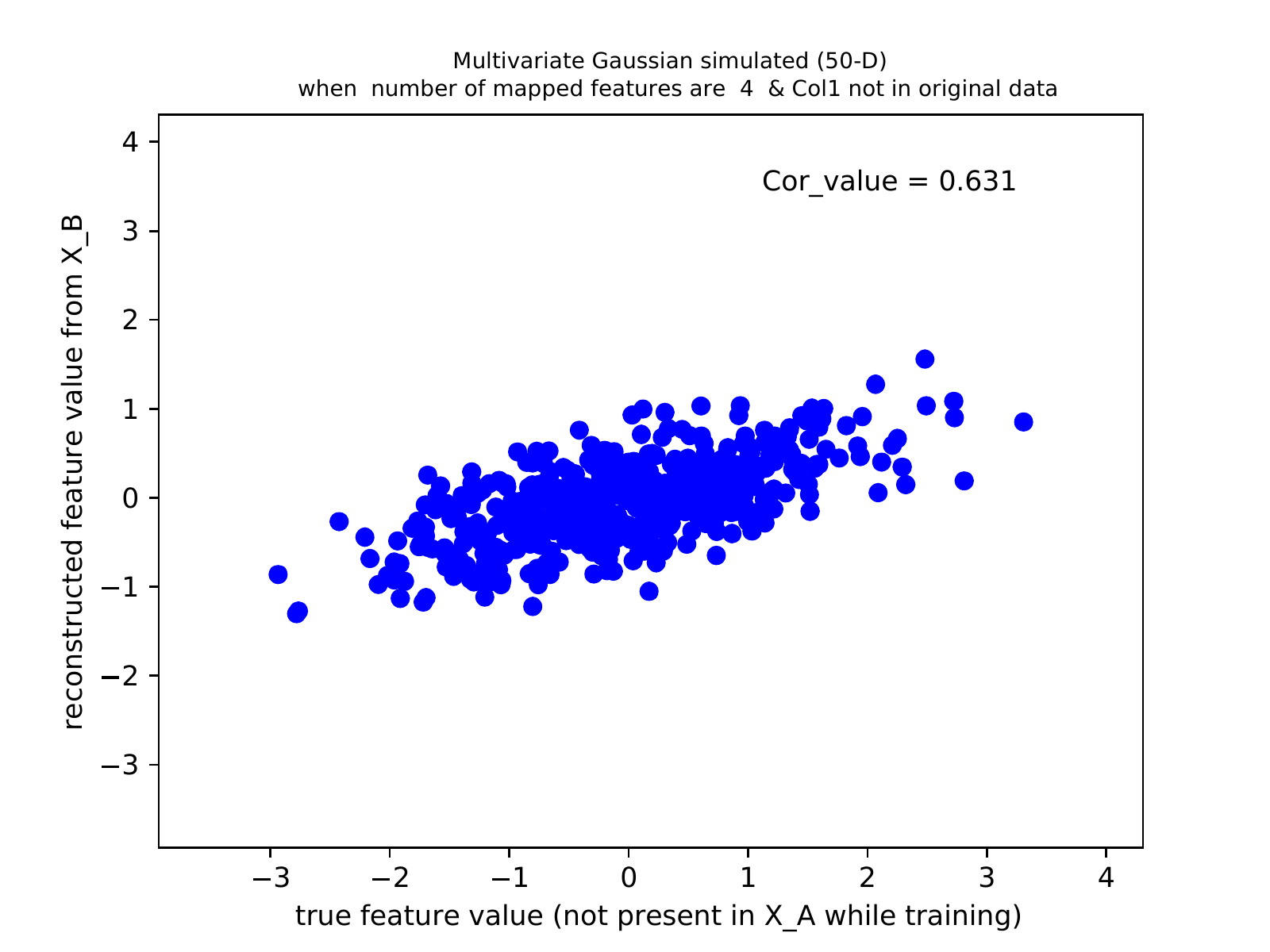}
        \caption{\footnotesize{}}
        \label{fig: Syn5_Chimeric_vs_increasing_number_of_features_inX2_Col1_surrogate}
    \end{subfigure}

    \vskip\baselineskip
    
    \begin{subfigure}[b]{0.475\textwidth}  
        \centering 
        \includegraphics[width=0.8\textwidth]{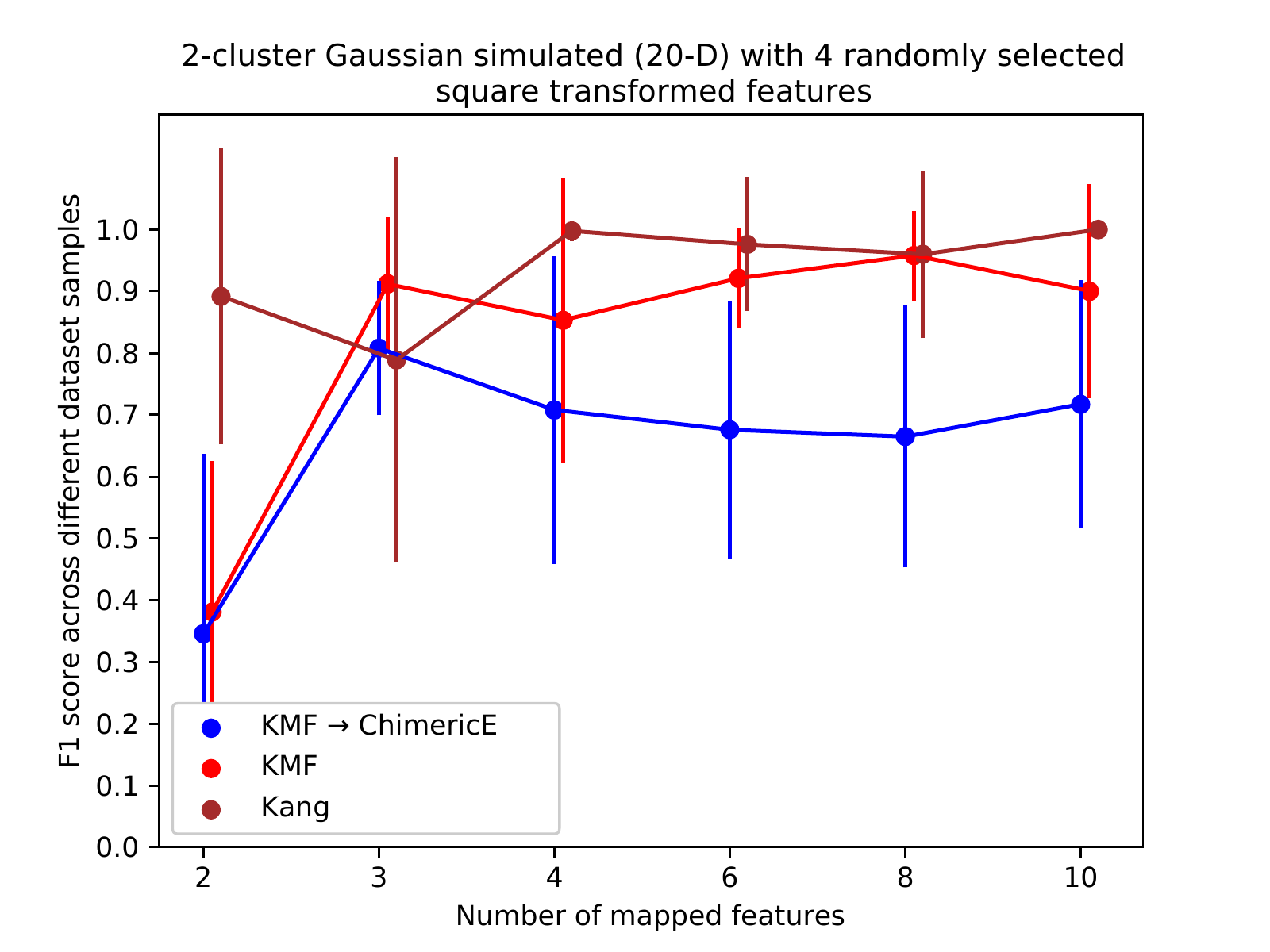}
        \caption{\footnotesize{}}
        % \caption{\footnotesize{\textbf{SD1}: Comparison wrt number of mapped features when some features had square transformation in $x^2$}}
        \label{fig: Syn1_Chimeric_vs_mappedfeatures_sq_transformed}
    \end{subfigure}
    \hfill
        \begin{subfigure}[b]{0.475\textwidth}  
        \centering 
        \includegraphics[width=0.8\textwidth]{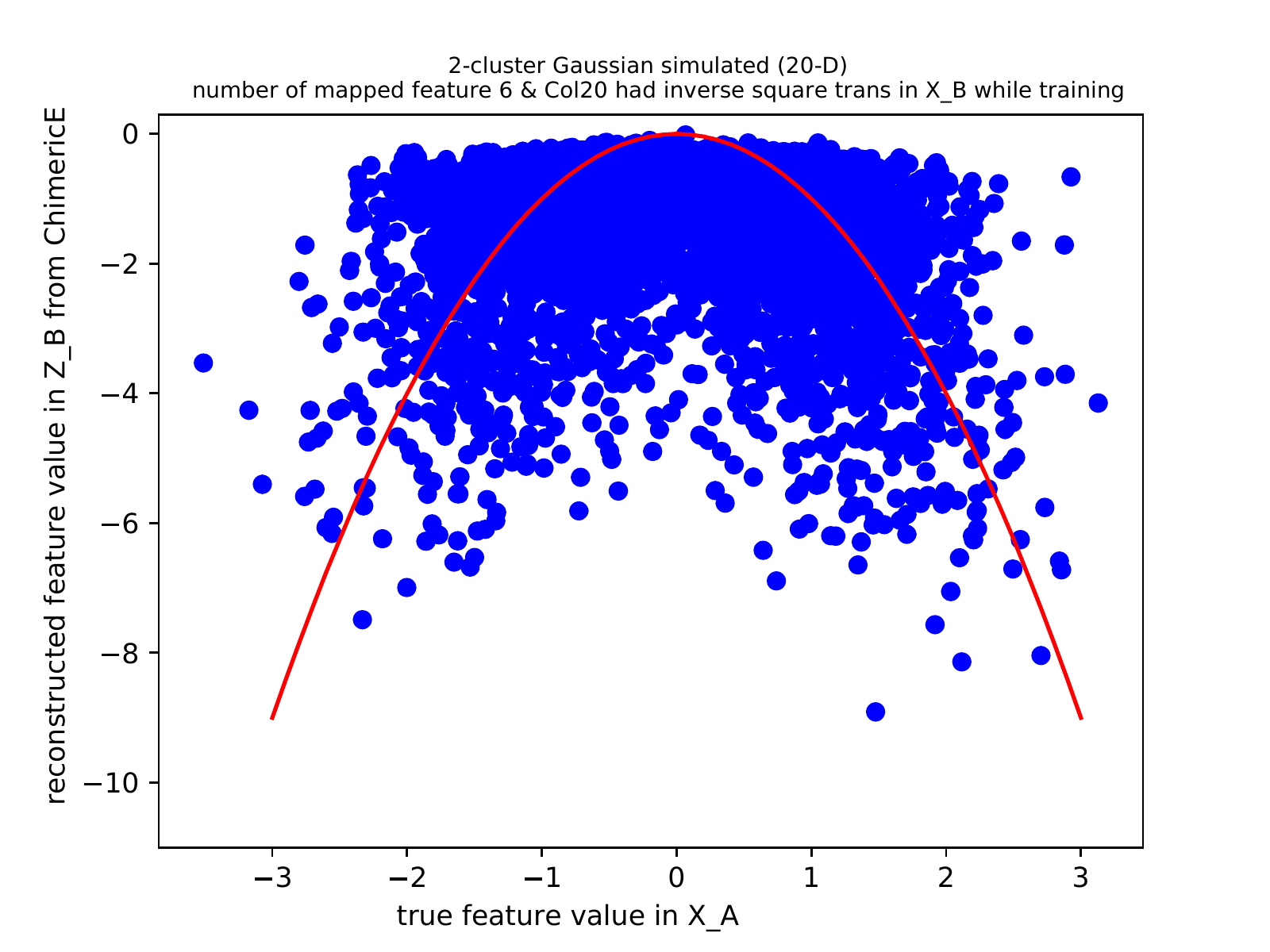}
        \caption{\footnotesize{}}
        % \caption{\footnotesize{\textbf{SD1}: Comparison wrt number of mapped features when some features had square transformation in $x^2$}}
        \label{fig: Syn1_Chimeric_vs_mappedfeatures_Invsq_transformed_Recons_Col20}
    \end{subfigure}

   \caption{Average F1 score for the KMF initialized chimeric encoder based schema matching in comparison to Kang method and KMF method on synthetic datasets. \textbf{2-cluster Gaussian simulated (20-D)}: (a) $x^A$ and $x^B$ have equal number of columns but they are permuted,  (b) both $x^A$ and $x^B$ have been binarized by thresholding all values at 0 after the original dataset has been normalized.
   \textbf{Multivariate Gaussian simulated (50-D)}: (c) The plot shows the performance comparison when $x^A$ has 15 features and there is a partial map between the features of $x^A$ and $x^B$ (with 4 pre-mapped features) (d) an illustration of chimeric encoder being able to learn a feature's representation (not present while training) using the surrogate correlations. \textbf{2-cluster Gaussian simulated (20-D)}: (e) $x^A$ and $x^B$ have equal number of columns but $5$ random features in unmapped set in $x^B$ were square transformed, (f) an illustration of chimeric encoder learning an inverse square transformation when 6 features were pre-mapped. }
   \label{mixture_fig}
\end{figure}

Figure \ref{mixture_fig} shows the performance on the somewhat more complex mixture distribution synthetic data.
%of two multivariate Gaussians, transformation of some features in $x^A$, and discrete data.
Figure \ref{fig: Syn1_Chimeric_vs_mappedfeatures} shows that KMF, chimeric, and Kang methods are all successful with mixture data. 
The F1 score is little smaller when these features are transformed to binary data (Figure \ref{fig: Syn1_Chimeric_vs_mappedfeatures_binarized}), but again all three methods work well.
Figure \ref{fig: Syn1_Chimeric_vs_mappedfeatures_sq_transformed} shows that transforming variables overall degrades performance. 
As expected, KMF and chimeric (which use Pearson correlation in this setup) are more affected than Kang, which is using mutual information as its dependence measure and should therefore be minimally affected by transformations.
Figure \ref{fig: Syn1_Chimeric_vs_mappedfeatures_Invsq_transformed_Recons_Col20} presents an example of chimeric encoder learning an inverse square transformation for a feature from $x^A$. We attempted to generate a similar plot for RadialGAN to set as a baseline. However, RadialGAN was unable to learn the transformation for similar settings (as shown in Figure \ref{fig: RG_not_learning_inv_trans} in Appendix \ref{app: no-match_FDR_examples}).
%Finally, we evaluate the chimeric encoder on a more realistic challenge: the partial mapping case. 

Figure \ref{fig: Syn5_Chimeric_vs_increasing_number_of_features_inX2_PM} is the more challenging setup of incomplete overlap (columns randomly dropped from both datasets).
We vary the total number of features in $x^B$ with the number of pre-mapped features fixed at 4.
While for all methods there is a decrease in F1 score values, the KMF and chimeric encoder methods are substantially better than the baseline. %
Figure \ref{fig: Syn5_Chimeric_vs_increasing_number_of_features_inX2_Col1_surrogate} displays the unobserved true values on an unshared feature and the reconstruction from chimeric encoder, showing the method's ability to learn to impute unobserved features from those that act as surrogate. 

We observed that the Gale-Shapley matching step's performance was much worse when reversing the direction of application (large database applying to small database). This was evident as the difference between the size of feature set of $x^A$ and $x^B$ increases, the gap between the performance of two directions also increases. An illustration of this phenomenon is presented in Figure \ref{fig: REal_Data_partial_mapping_5_extra_two_Stage_from_CCx1_vsCCx2} in Appendix \ref{app: no-match_FDR_examples}. 
Our initial experiments also suggested that the choice of the encoder output dimension in the chimeric method has a significant effect on its accuracy. An example for \textbf{2-cluster Gaussian simulated (20-D)} with dim$=5$ and dim$=10$ is provided in Figure \ref{fig: SD1_compar_L} of Appendix \ref{app: L_dim_comp}  and clearly, as the dimension decreases, the chimeric AE gets better at mapping between databases. This affirms that our algorithm relies on the compression power of the autoencoders. 
%We further increase the compression by using dropout.

% Figure \ref{fig: Syn5_Chimeric_vs_increasing_number_of_features_inX2} shows an example of case (2) for \textbf{SD5} where $x^1$ had $15$ features and the number features are increased in $x^2$; all features in $x^1$ need not have a match in $x^2$ (\textcolor{red}{to update after the experiment}). 
% to get Figure \ref{fig: Syn6_Chimeric_vs_increasing_number_of_features_inX2}.

%In all the above figures using synthetic data, we generated 5 datasets, subsampled 4 times, and selected 3 permutations.

% \textbf{Choice of sample size on the performance}

% \begin{figure*}[h!]
%     \centering
%     \begin{subfigure}[b]{0.475\textwidth}
%         \centering
%         \includegraphics[width=1.1\textwidth]{Figures_final/All_matchesASym_Chim_Syn_6_varyingData_num_total_fea_inX2_5.pdf}
%         \caption{\footnotesize{}}
%         \label{fig: Syn6_Chimeric_vs_increasing_number_of_features_inX2}
%     \end{subfigure}
%     \hfill
%     \begin{subfigure}[b]{0.4475\textwidth}  
%         \centering 
%         \includegraphics[width=0.4\textwidth]{Figures_final/Not_All_matchesASym_Chim_Syn_6_varyingData_num_total_fea_inX2_6.pdf}
%         % \caption{\footnotesize{\textbf{SD1}: Comparison wrt number of mapped features when some features had square transformation in $x^2$}}
%         \caption{\footnotesize{}}
%         \label{fig: Syn5_Chimeric_vs_increasing_number_of_features_inX2}
%     \end{subfigure}
%   \caption{Incomplete overlap between features of $x^1$ and $x^2$. (a) \textbf{SD6:} The plot shows the performance of Chimeric approach when $x^1$ has 10 features and feature set of $x^2$ even though larger, is a superset for the former. (b) \textbf{SD5:}}
% \end{figure*}

\begin{table}[]
{\scriptsize
\centering
\begin{tabular}{|l|c|c|c|}
\hline
\multicolumn{1}{|c|}{\textbf{Dataset  and scenarios}}                                                                                                        & \textbf{Kang}  & \textbf{KMF}   & \textbf{KMF $\rightarrow$ ChimericE} \\ \hline
\begin{tabular}[c]{@{}l@{}}Multivariate Gaussian simulated (20-D):\\  Avg F1 score vs mapped features (Perm)\end{tabular}                                    & 0.782          & \textbf{0.937} & 0.924                                 \\ \hline
\begin{tabular}[c]{@{}l@{}}Multivariate Gaussian simulated (20-D): \\ Avg F1 score vs sample size (Perm)\end{tabular}                                        & 0.388          & \textbf{0.919} & 0.728                                 \\ \hline
\begin{tabular}[c]{@{}l@{}}Multivariate Gaussian simulated (20-D): Avg F1 score \\ vs number of features in larger database (Onto)\end{tabular}              & 0.715          & 0.773          & \textbf{0.808}                        \\ \hline
\begin{tabular}[c]{@{}l@{}}2-cluster Gaussian simulated (20-D): \\ Avg F1 score vs mapped features (Perm)\end{tabular}                                       & 0.949          & 0.934          & 0.932                                 \\ \hline
\begin{tabular}[c]{@{}l@{}}Binarized 2-cluster Gaussian simulated (20-D): \\ Avg F1 score vs mapped features (Perm)\end{tabular}                             & \textbf{0.977} & 0.918          & 0.905                                 \\ \hline
\begin{tabular}[c]{@{}l@{}}2-cluster Gaussian simulated (20-D) with some non-linear \\ transf features : Avg F1 score vs mapped features (Perm)\end{tabular} & \textbf{0.936} & 0.819          & 0.652                                 \\ \hline
\begin{tabular}[c]{@{}l@{}}Multivariate Gaussian simulated (50-D): Avg F1 score \\ vs number of features in larger database (Partial mapping)\end{tabular}   & 0.393          & \textbf{0.689} & 0.646                                 \\ \hline
ACTFAST data: Avg F1 score vs mapped features (Perm)                                                                                                         & 0.282          & \textbf{0.961} & 0.702                                 \\ \hline
ACTFAST data: Avg F1 score vs  sample size (Perm)                                                                                                            & 0.523          & \textbf{0.770} & 0.472                                 \\ \hline
\begin{tabular}[c]{@{}l@{}}ACTFAST data: Avg F1 score vs number of features in \\ larger database (Onto)\end{tabular}                                        & \textbf{0.943} & 0.878          & 0.882                                 \\ \hline
\begin{tabular}[c]{@{}l@{}}ACTFAST data: Avg F1 score vs number of features in \\ larger database (Partial mapping)\end{tabular}                             & \textbf{0.861} & 0.832          & 0.805                                 \\ \hline
MIMIC data: Avg F1 score vs mapped features (Partial mapping)                                                                                               &         0.148        &         \textbf{0.484}       &         0.424                              \\ \hline
\end{tabular}}
\caption{Table showing the average F1 score for different methods in different scenarios. Bold values denote the method with a statistically significant higher average F1 score in comparison to others based on the Wilcoxon test.}
\label{tab:significance_test_results}
\end{table}

\vspace{2cm}
\subsection{Real data results} \label{sec: real_data_experiments}

% In this section, we present illustrative experiments on the real ACTFAST dataset. 

\paragraph{\textbf{MIMIC-III}}
%The MIMIC-III dataset provides a good reflection of how complex the schema matching problem can be in the real world. Some prominent issues are as follows. 
% it presents a realistic picture of how severe the schema matching problem could be. 
%% why is the # mapped correctly higher for chimeric when the f1 is lower?
% On an average, the proposed algorithms are able to correctly match 27 out of 49 (KMF) and 29 out of (KMF initialized Chimeric) when 58 features are known-mapped even in this difficult setting as  

\begin{figure}
    \centering
    \begin{subfigure}[b]{0.478\textwidth}
        \centering
        \includegraphics[width=1.0\textwidth]{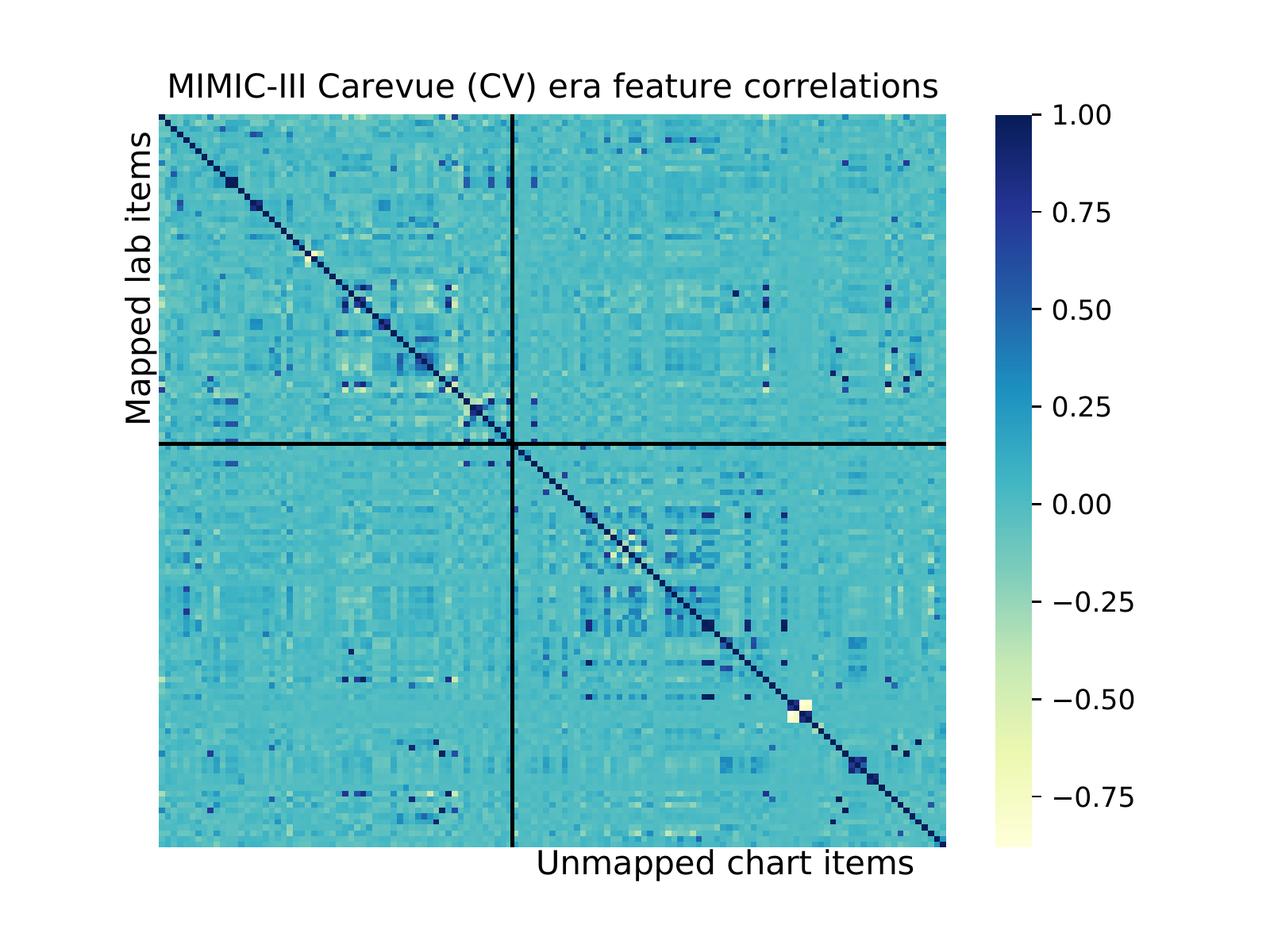}
        \caption{\footnotesize{}}
        % {}    
        \label{fig: MIMIC_correlation_clustered_CV}
    \end{subfigure}
    \hfill
    \begin{subfigure}[b]{0.478\textwidth}  
        \centering 
        \includegraphics[width=1.0\textwidth]{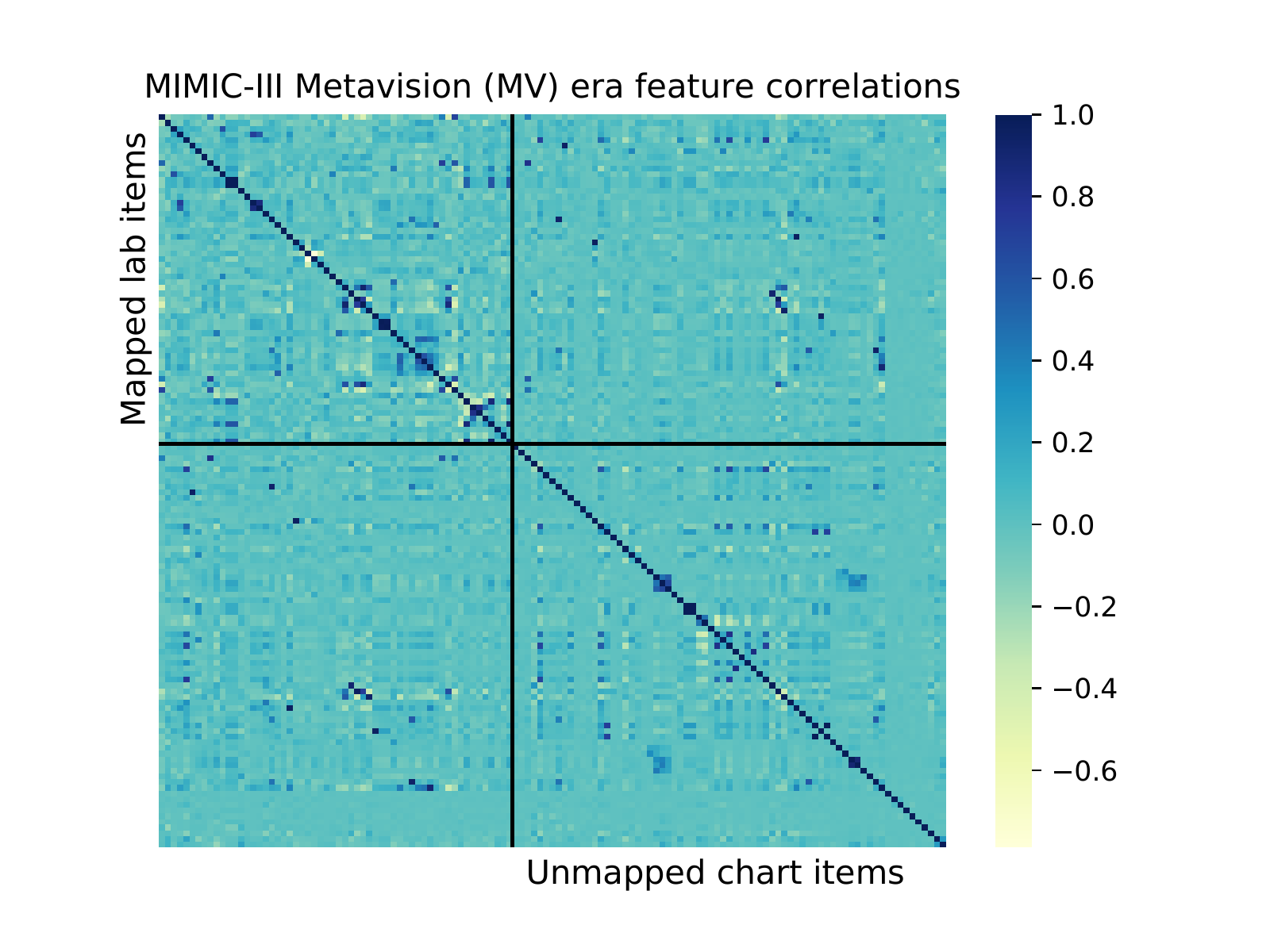}
        \caption{\footnotesize{}}
        % {}       
        \label{fig: MIMIC_correlation_clustered_MV}
    \end{subfigure}
    \vskip\baselineskip
        \begin{subfigure}[b]{0.478\textwidth}
        \centering
        \includegraphics[width=1.0\textwidth]{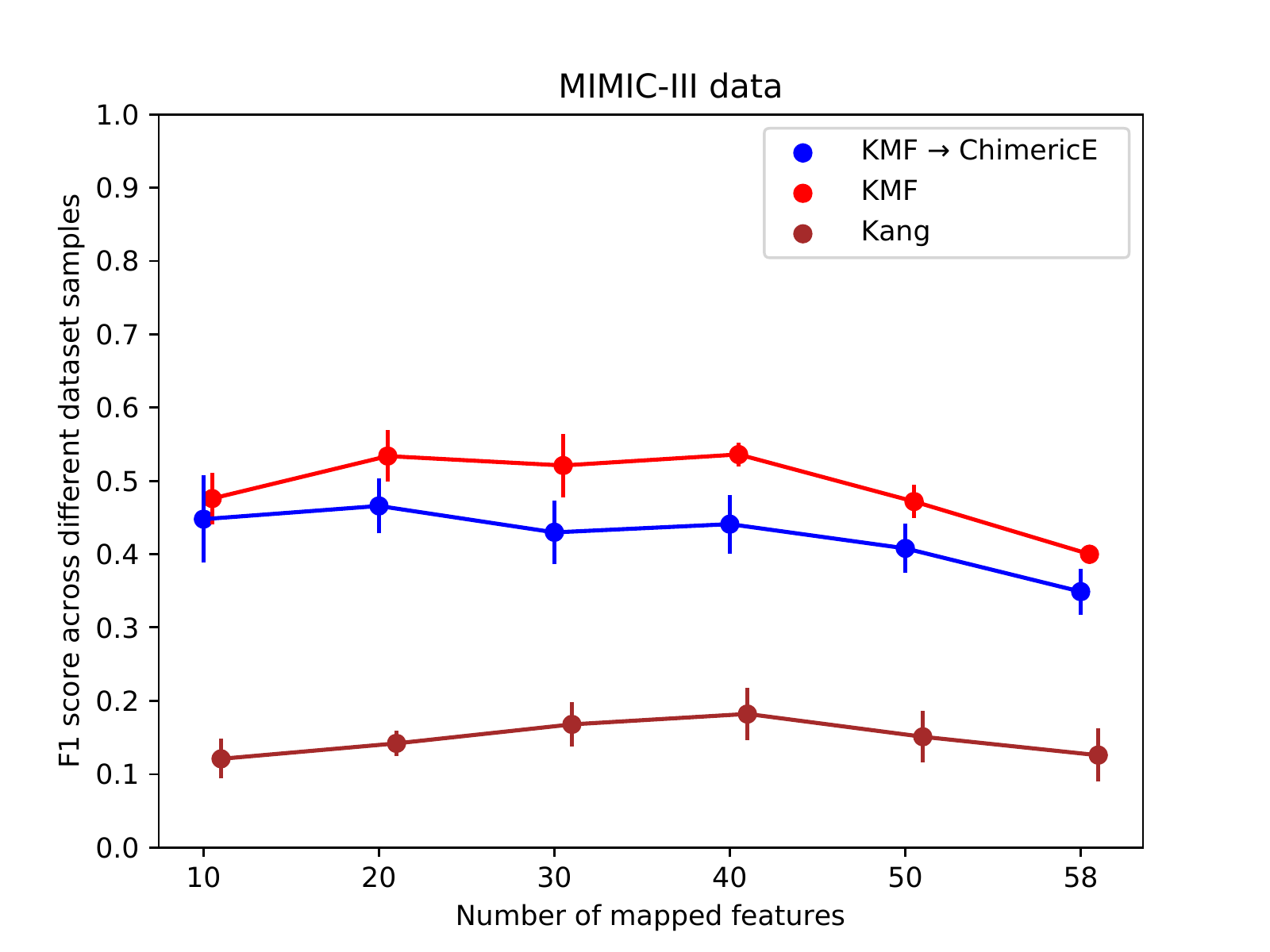}
        \caption{\footnotesize{}}
        % {}    
        \label{fig: MIMIC_Chimeric_vs_mapped_features_PM}
    \end{subfigure}
    \hfill
    % \begin{subfigure}[b]{0.478\textwidth}  
    %     \centering 
    %     \includegraphics[width=0.8\textwidth]{}
    %     \caption{\footnotesize{}}
    %     % {}       
    %     \label{fig:}
    % \end{subfigure}
    \vskip\baselineskip
    \caption{Performance of chimeric encoder in terms of average F1 score on MIMIC-III dataset. Correlation structure within features (labevents and chartevents combined) of MIMIC-III data in (a) Carevue (CV) era and (b) Metavision (MV) era. (c) Average F1 score for schema matching between CV and MV era data by chimeric encoder and baselines.}
\end{figure}

Figure \ref{fig: MIMIC_correlation_clustered_CV} and \ref{fig: MIMIC_correlation_clustered_MV} show the correlograms for the Carevue and Metavision eras respectively.
Although the number of mapped features is fairly high, the correlation between individual mapped and unmapped features is generally low.
%The labevent item and chartevent item with highest correlation in CV era is 'Calculated Total CO2' and 'Arterial CO2 Pressure'.
%The lab item and chartevent item with highest correlation in MV era is 'Base Excess' and 'Arterial CO2 Calc'.
Figure \ref{fig: MIMIC_Chimeric_vs_mapped_features_PM} shows the matching performance while varying the number of included mapped laboratory features.
The performance is overall lower (correctly mapping at most 22 of 49 features), as we might expect given the weak dependence between features and the large fraction of features not shared between the two datasets.
Even after choosing the optimal parameters for the baseline method (Kang), its performance is extremely poor compared to our KMF and chimeric encoder methods.
%The proposed algorithms are able to match almost $50\%$ correctly even in this difficult setting as can be seen in Figure . 
%As can be observed in the plot, the proposed methods perform better than the baseline Kang method. 
%We would like to note that we separately tuned for the parameter $alpha$ used in Kang method. 
%
The cross-validation hyperparameter tuning strategy when applied to the MIMIC dataset yielded an average F1 score of $0.40$, which was modestly lower than the global optimum of $0.45$.
%% Are these in a meaningful order? You shouldn't expect the reader to refer back to another section to guess.
% The above steps were repeated for each hyperparameter tuple and the hyperparameter tuple with largest average F1 score ($0.975862$) across 10 trials is selected as the best hyperparameter $(1.0, 1.4, 32, 0.5, 40, 0.001)$.
% We observed that the hyperparameter tuple $ (0.4, 1.0, 64, 0.7, 20, 0.01)$ has the highest F1 score ($0.4528$) on $49$ unknown chartevent features.
% The corresponding F1 score on $49$ chartevent features obtained by using the tuple $(1.0, 1.4, 32, 0.5, 40, 0.001)$ from  above proposed hyperparameter tuning strategy is $0.4$ which is close to what we obtained by exhaustive search.
On an average, the proposed algorithms are able to correctly match 22 out of 49 (KMF) and 20 out of 49 (KMF initialized Chimeric) unknown chartevents features when 58 features are known-mapped.

\begin{figure}
    \centering
    \begin{subfigure}[b]{0.478\textwidth}
        \centering
        \includegraphics[width=0.8\textwidth]{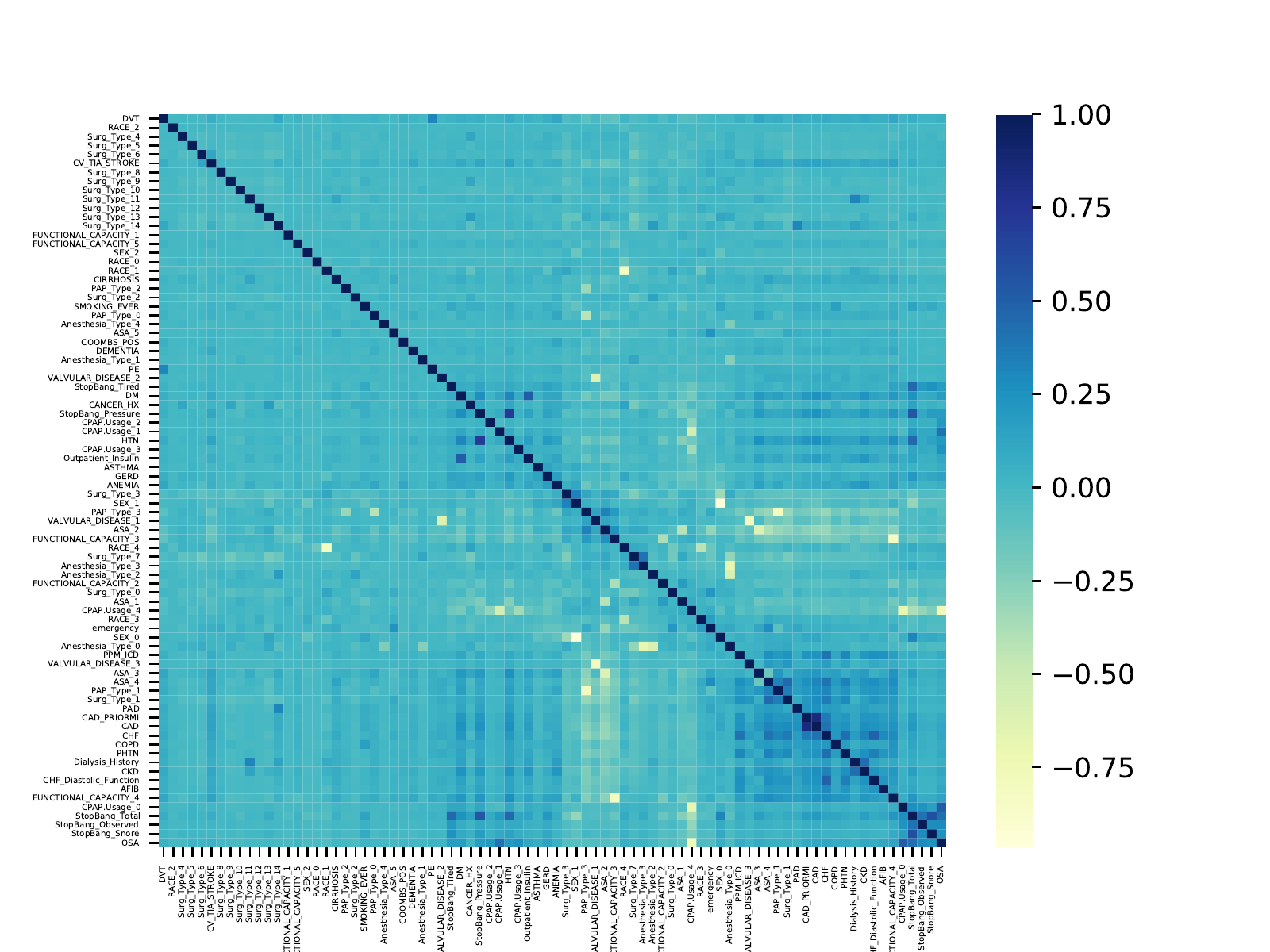}
        \caption{\footnotesize{}}
        % {}    
        \label{fig: Real_correlation_clustered}
    \end{subfigure}
    \hfill
    \begin{subfigure}[b]{0.478\textwidth}  
        \centering 
        \includegraphics[width=0.8\textwidth]{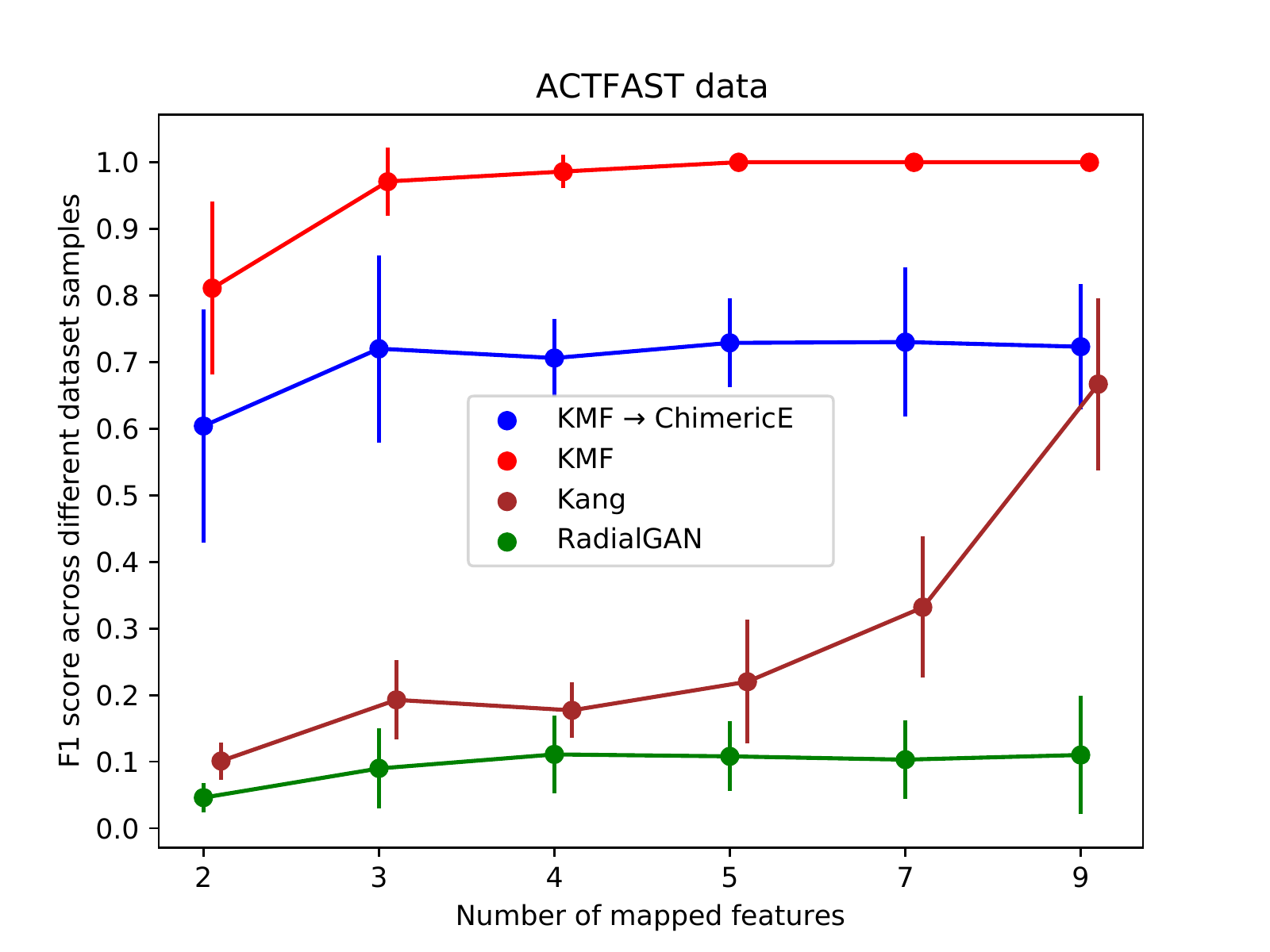}
        \caption{\footnotesize{}}
        % {}       
        \label{fig: Real_comp_Radial_GAN_Chimeric_vs_mappedfeatures}
    \end{subfigure}
    \vskip\baselineskip
        \begin{subfigure}[b]{0.478\textwidth}
        \centering
        \includegraphics[width=0.8\textwidth]{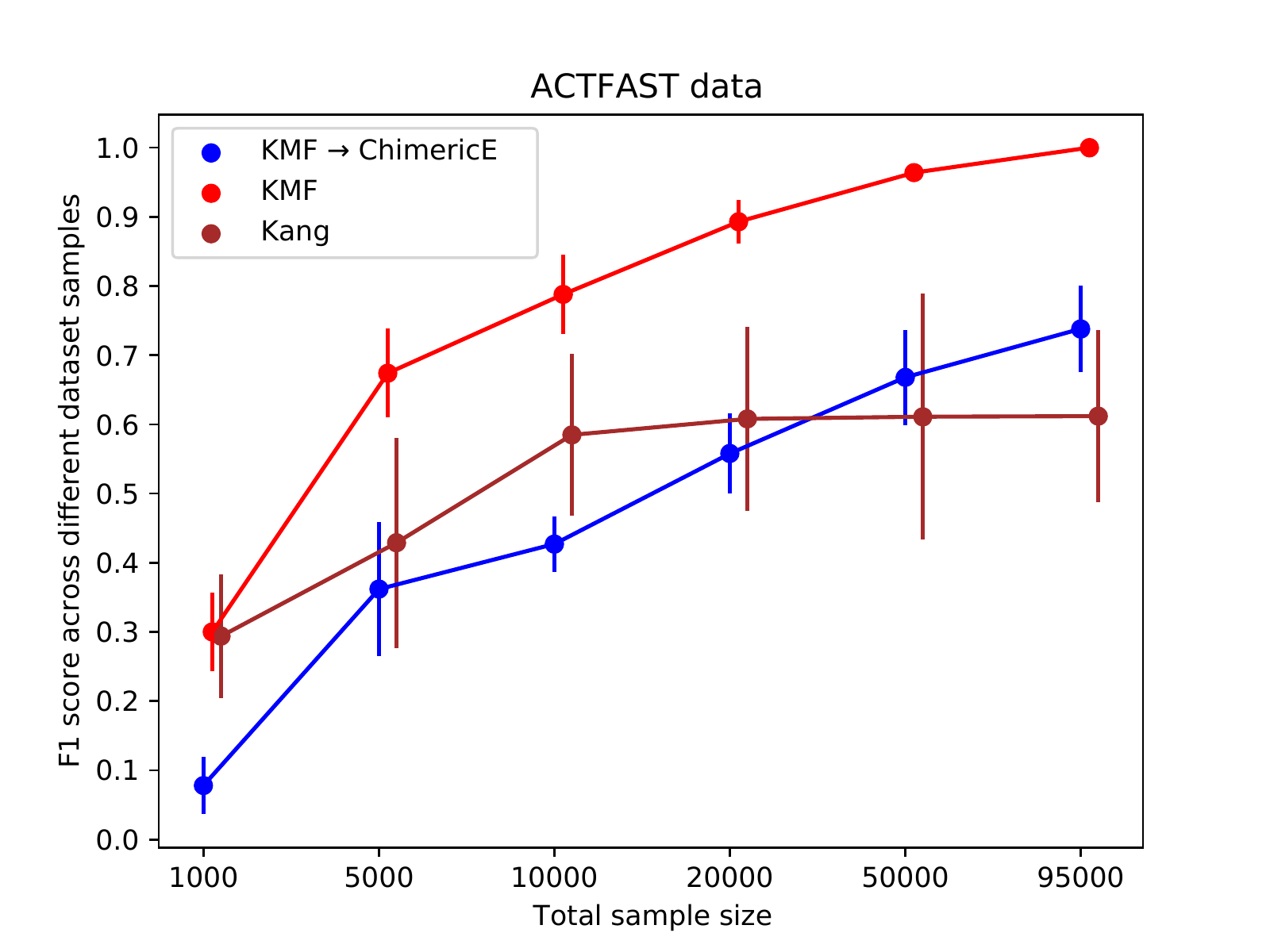}
        \caption{\footnotesize{}}
        % {}    
        \label{fig: Real_Chimeric_vs_sample_size}
    \end{subfigure}
    \hfill
    \begin{subfigure}[b]{0.478\textwidth}  
        \centering 
        \includegraphics[width=0.8\textwidth]{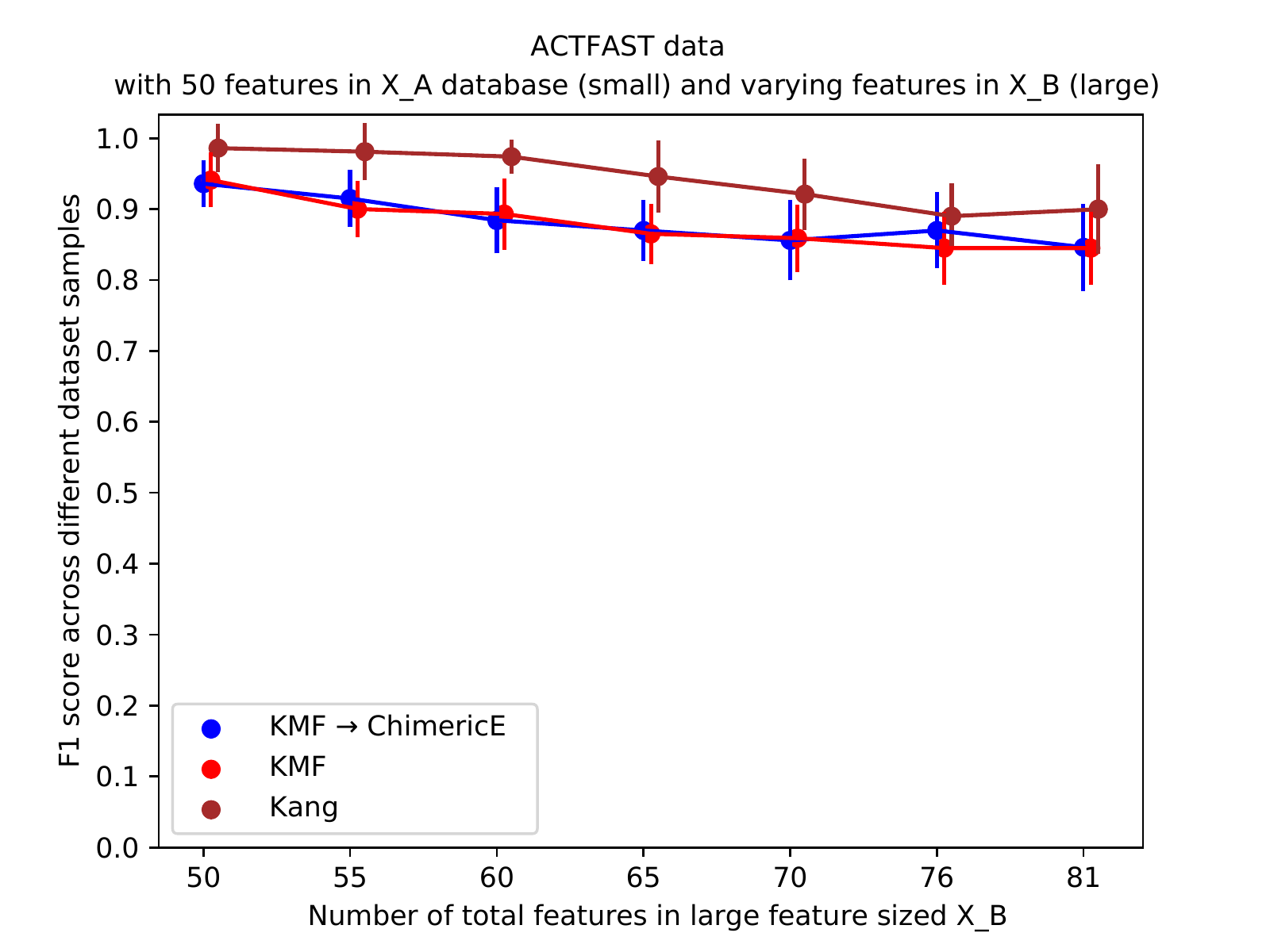}
        \caption{\footnotesize{}}
        % {}       
        \label{fig: Real_Chimeric_vs_increasing_number_of_features_inX2}
    \end{subfigure}
    \vskip\baselineskip
    \begin{subfigure}[b]{0.478\textwidth}
        \centering
        \includegraphics[width=0.8\textwidth]{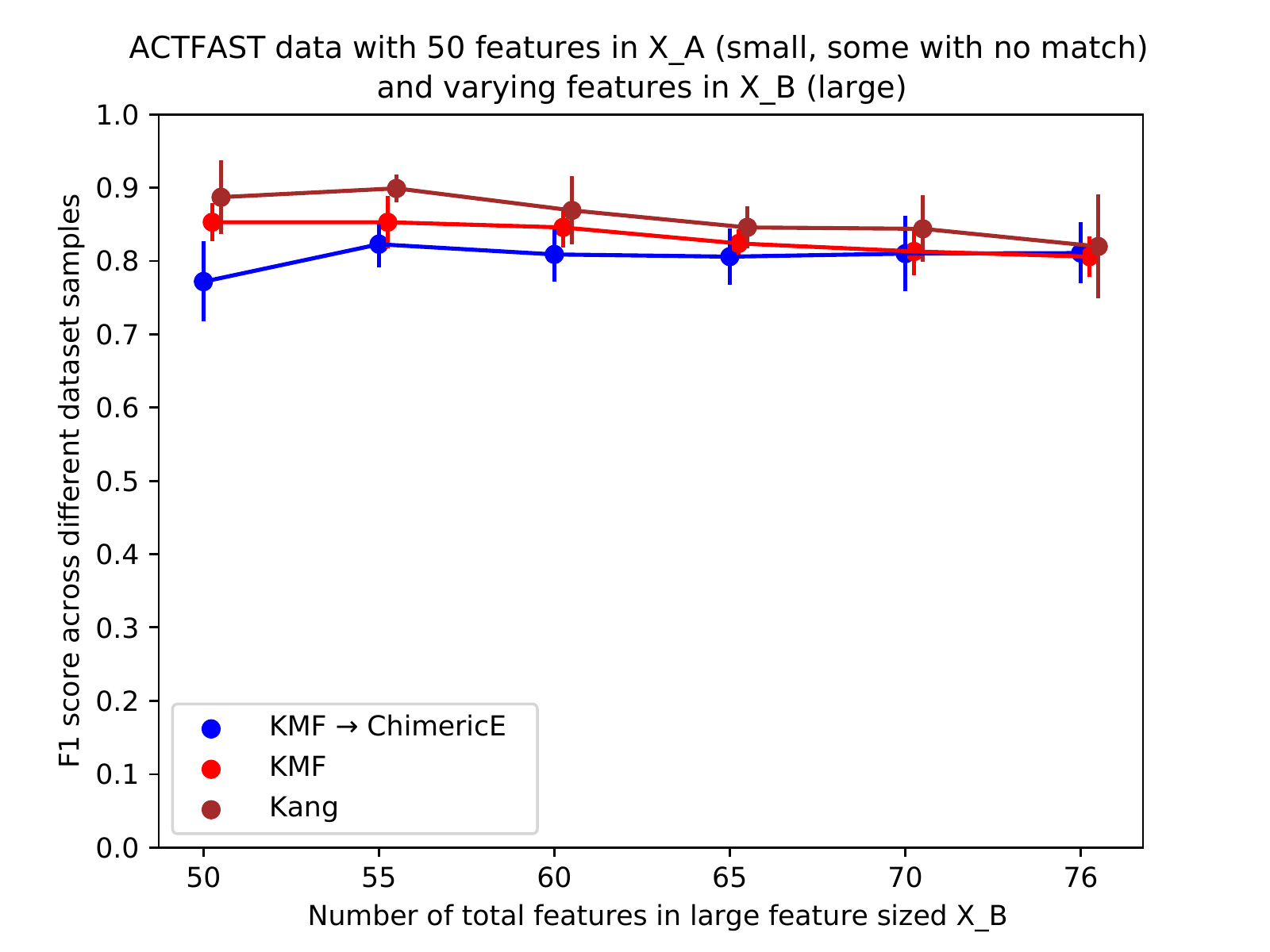}
        \caption{\footnotesize{}}
        % {}    
        \label{fig: Real_Chimeric_vs_increasing_number_of_features_inX2_not_all_matches}
    \end{subfigure}
    \hfill
    \begin{subfigure}[b]{0.478\textwidth}  
        \centering 
        \includegraphics[width=0.8\textwidth]{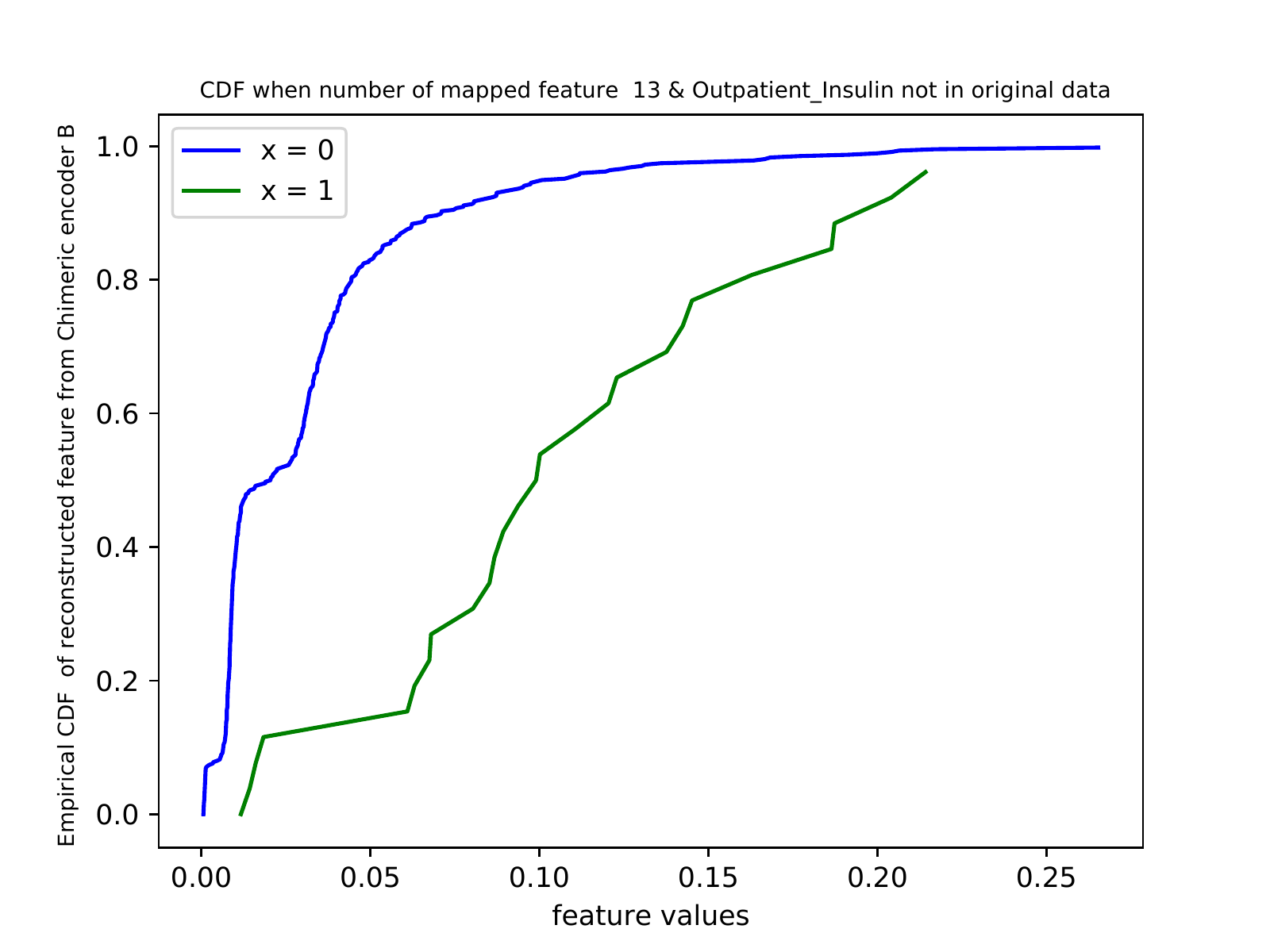}
        \caption{\footnotesize{}}
        % {}       
        \label{fig: Real_data_no_true_match_surrogate_outpatient_insulin}
    \end{subfigure}
    \caption{Performance of chimeric encoder in terms of average F1 score on ACTFAST dataset (a) Correlation structure within features (after one-hot encoding) of ACTFAST data obtained after spectral clustering with 10 clusters (b) Comparison between proposed methods and the baselines (c) Performance variation with increase in the total sample size when number of mapped features were fixed at 5 categorical variables (d) Effect of feature overlap extent on F1 score when the consortium size is only 50 features (onto map case, 3 pre-mapped features) (e) Effect of feature overlap extent on F1 score when the consortium size is only 50 features (partial map case, 3 pre-mapped features) (f) Empirical cumulative distribution function of reconstructed binary feature using chimeric encoder that had `no-match'; difference in two curves show that the reconstructed feature values ($\{0,1\}$) are well discriminated.}
\end{figure}

\paragraph{\textbf{ACTFAST}}
Figure \ref{fig: Real_correlation_clustered} presents a correlogram for the ACTFAST dataset.
Some features like ASA, Anesthesia type, and Functional capacity are substantially correlated with many variables.
%However, there are others like surgery types and sex that are weakly correlated with other binary variables. 
%
In Figure \ref{fig: Real_comp_Radial_GAN_Chimeric_vs_mappedfeatures}, we present F1 score while varying the number of mapped features, and in Figure \ref{fig: Real_Chimeric_vs_sample_size} we vary the sample size.
KMF is consistently the best performing algorithm, and RadialGAN is consistently the worst.
Chimeric encoding is superior to the Kang baseline when a small number of features are mapped, and the two are largely equivalent otherwise.
As we saw before, the chimeric encoder is more sample-size dependent than any other method because of its complexity.
Although the sample size of ACTFAST dataset is larger than the other setting we consider, the frequency of the binary features is much less, and as a result the performance has not saturated with the entire dataset used.
The large standard errors are a function of the random selection of variables to treat as pre-mapped, since a few variables are very informative. 

Figure \ref{fig: Real_Chimeric_vs_increasing_number_of_features_inX2} displays mapping accuracy as a function of the number of unshared features in the case where one column space is a subset of the other, and Figure \ref{fig: Real_Chimeric_vs_increasing_number_of_features_inX2_not_all_matches} displays the same when two column spaces have unique features too.
When the number of features in the consortium is fixed at $50$ and the size of the new database interested in joining is increased, we observe that the F1 score does not show a substantial decrease. 
The Kang baseline has a statistically significant edge in performance averaged over settings in this experiment (Table \ref{tab:significance_test_results}), but the absolute difference performance is small with all 3 methods performing about the same. 
The findings are similar in the partial mapping experiment in Figure \ref{fig: Real_Chimeric_vs_increasing_number_of_features_inX2_not_all_matches}.
We investigated the lower F1 scores in the chimeric and KMF methods, 
% and found that the step-down procedure used to control the match-acceptance threshold.
and found that the chimeric and KMF methods were declaring incorrect matches to exist when a statistically significant surrogate was identified.
Given that some of the clinical entities in the assessment are closely correlated (seen in Figure \ref{fig: Real_correlation_clustered}), we feel that this represents more of a problem with the evaluation metric. 
Figure \ref{fig: Real_data_no_true_match_GS_matches_correlation_SD_acceptes_fromX1} in Appendix \ref{app: no-match_FDR_examples} depicts a histogram of density of the true correlations on the GS matches for `no-match' features that were declared significant by step down procedure.
Most of these mistakes are due to the `no-match' features being matched to correlated surrogates as seen on the right half of the plot.
% We provide some example of the GS matched feature pairs, their frequency and the corresponding correlation in Table \ref{tab: no-match_mistakes_correlation} in Appendix \ref{app: no-match_FDR_examples}.

However, the surrogate feature matches have the advantage of allowing one to create composite measures or reconstruct features that are not present in one dataset using the chimeric encoder. 
Figure \ref{fig: Real_data_no_true_match_surrogate_outpatient_insulin} shows such an example, displaying the cumulative distribution function of the reconstruction of a binary feature ``Outpatient insulin'' that was not present in $x^A$, stratified based on the true unobserved value. 
The distribution of this feature is very skewed with only $0.052$ positive ratio. 
The distribution functions for the two populations are well separated, which is an evidence for good reconstruction of a binary feature at a variety of potential thresholds.

\section{Discussion} \label{sec: discussion}

The problem of combining two data sources with shared features is complex and multi-step in applied work.
% Combining two data sources with shared features is a complex and multi-step problem in implementation science.
Recently, major software firms have put forward proposals to match healthcare databases \footnote{1) \url{https://cloud.google.com/healthcare} 2) \url{https://aws.amazon.com/healthlake/} }.
Most approaches to this task either rely on extensive meta-data or columns with very distinct distributions, such as text fields.
We envision our contribution as addressing a specific step in these applications.
After identifying easily mapped features using the above sources of information, 
we focus on the challenging case of binary or continuous features without substantial metadata.
We further assume that variable shift or rescaling in continuous variables make recognizing univariate distributions unreliable.
In that setting, we proposed a procedure that jointly learns to match features between databases, identify transformed features, and impute missing features when a surrogate is present.

Our method has several novel contributions.
First, our use of Gale-Shapley matching provides a locally optimal transformation of a similarity matrix to a set of mappings.
The procedure is fast compared to the heuristic stochastic search proposal of Kang and colleagues \cite{kang_schema_2008} and provides candidates for the next-best matches, should the reviewer find a proposal to be incorrect.
In practice, we would expect the final review stage of matching the two databases to consider a small number of mappings, so the incorrect modifications by the chimeric encoder are of little applied consequence.
Because Gale-Shapley is very fast, if any corrections are made by users it can be interactively re-run with those corrections enforced.
Gale-Shapley does have some disadvantages. 
First, it does not identify global optima, only a local one.
Second, the results depend on the selection of proposal direction.
Third, Gale-Shapley on its own does not produce any measure of uncertainty and requires an ad-hoc secondary step to identify when to stop accepting matches.
The procedure of matching between inputs and chimeric outputs can be seen as a form of knowledge distillation;
we are searching over the space of permutations (and discarding some features) for a function that approximates the chimeric encoder.
Search strategies built into the neural network optimization are certainly possible, but we defer this to future work.

Second, we are the first, to our knowledge, to use false discovery rate control to define a cutoff for partial mapping.
In our experiments with low dimensional latent classes or spaces, we find that this cutoff is generally liberal (allows too many matches) because a surrogate variable related to the same latent class is identified with low but statistically non-zero correlation.
By ranking the proposed matches by the estimated correlation, users can detect when the chimeric correlation is no longer meaningful.
As discussed below, a leave-k-out or cross-validation approach to validation is also possible, which can help define a threshold.

We are also the first to combine the schema matching problem with learning transformations, including multivariable transformations used to impute missing features.
%If the transformation is not one-to-one, it also changes the MI between datasets, impairing baselines which depend on this measure.
The chimeric encoder reveals the transformation both graphically and quantitatively with the highly-ranked matches.
We regard identifying \textit{semantically identical} features as a distinct task from transformed features, and recommend this as a multi-stage procedure. 
We used Pearson correlation to generate a similarity matrix between features using the chimeric data, which preferentially matches native versus transformed features.
As discussed above, this aspect of the proposal (linking a feature to a surrogate) probably explains its lower than expected F1 score in the partial mapping case.
Non-parametric associations like distance correlation \cite{szekely_measuring_2007} or mutual information expand the notion of ``matching'' to any one-to-one univariate transformation, which would modify the problem definition for KMF.
Multiple association measures per pair of features (for example, if the conditional relationship is not homogeneous and there is substantial distributional shift) are also possible.

Our work is naturally compared to that of Kang \cite{kang_schema_2003} and others following the same theme.
Kang relies upon the association between features to form a fingerprint and searches the space of permutations between databases to optimize the overall similarity of association matrices.
Interestingly, our simple initialization procedure performed better in almost all cases.
In the Gaussian case, the pairwise covariance matrix completely defines the dependency between features, and the method's poor performance was surprising.
We observed that in datasets with binary features the Kang method retains good performance.
%However, in our example with strongly non-Gaussian dependence the Kang methods performance was comparatively weak.
We identified several relative weaknesses of the Kang method.
First, it was sensitive to the choice of ``normal metric'' versus ``Euclidean metric'' (data not shown).
%Second, it required a larger sample size than our method.
Second, it performed poorly in the more realistic and difficult partial mapping case even after tuning over its free parameter.
Third, we observed it to have occasional dramatic failures resulting in wide standard deviations, presumably related to difficulty with its stochastic search over permutations.
One advantage of the Kang method is that it can work without any prior knowledge matching features (or a very small number); our method is engineered assuming this prior data is available.
In such cases, one can replace the correlation-fingerprint initialization we propose with Kang's method and rely on the chimeric encoder for transformation.

Our method also resembles the work of Yoon and colleagues in RadialGAN \cite{yoon_radialgan_2018}.
They have a similar approach to transformation between datasets, but they replace the known-mapped chimeric loss with a discriminator.
RadialGAN has no natural way to take advantage of prior information mapping some variables.
The RadialGAN authors explicitly do not attempt to match schemas, but there is no reason not to use their network to estimate similarity matrices across datasets.
We found RadialGAN to be conceptually problematic, because if there is feature shift between databases (such as different mixes of patient classes), then the algorithm will be forced to awkwardly transform patients from one class to another (otherwise the discriminator will easily notice the differences in distribution).
We note that our cycle consistency loss is slightly different from the same term in RadialGAN, $L(g^A(f^B(z^B)), x^A )$ versus $L(f^B(z^B), f^A(x^A) )$ where $z^B \equiv g^B(f^A(x^A))$. 
The RadialGAN cycle consistency loss shifts the comparison to the encoded latent space, which has artificial neural network output in both terms.
During development, we found that this version of cycle consistency requirement leads to difficulty in training similar to mode collapse in other GANs.
That is, various undesirable solutions (like a constant value in the encoded space) maximize the discriminator loss.
Because our cycle consistency is anchored to real data, these mode collapse problems did not occur.
We found the RadialGAN method to suffer slow and inconsistent training leading to poor evaluation metrics despite multiple attempts to tune it for this purpose.
However, the inclusion of both cross-reconstruction and cycle consistency losses improved matching performance in the early development of the chimeric encoder approach, and we suspect cycle consistence is a valuable component encouraging the unmapped features to be effectively represented.
Unfortunately, the RadialGAN authors were not able to share code or data for us to further understand the differences between their experiments and ours.

\subsection{Limitations}

Our method assumes that the data has a true latent representation (effective compression) with a substantial association between the known-mapped features and unmapped features to allow the latent spaces to be aligned.
We found that the results were sensitive to the degree of compression in the encoder architecture.
For example, with too much compression the associations between features and their chimeric decoding are unreliable, making the matching stage struggle.
On the other hand, with too little compression there is no shared representation to align the latent spaces; in the limiting case, the encoder could simply learn an identity transformation.
The encoder hyperparameters like dropout are also important, as dropout encourages the encoder to spread out influence for the latent representation.
The addition of nonlinearity was also important; permutation is a linear operator, and so a network without nonlinearity can learn such a transformation effectively.
However, without nonlinearity, the encoder struggles with nonlinear representation in the latent space and nonlinearly transformed variables.
There was tension when choosing hyperparameters between the quality of the mapping and imputation (requires more compression) and quality of transformation reconstructions (favors less compression).
This suggests that a multistage procedure with distinct hyperparameters for the mapping and transformation stages can further improve this work.
We expect that users seeing output like in Figure \ref{fig: Syn1_Chimeric_vs_mappedfeatures_Invsq_transformed_Recons_Col20} would fit a polynomial or similar relationship to the $(x, z)$ pairs and update the data to include the transformation.
Although the chimeric encoder may lose a substantial amount of accuracy in reconstruction with heavy compression, we stress that this is unimportant if a mapping or transformation is ultimately discovered.

Selecting these hyperparameters is difficult.
As demonstrated for the MIMIC-III dataset, we can withhold some prior-mapped inputs from the algorithm and estimate its performance in matching them.
This is easy when the number of mapped features is large and they have saturated their contribution to learn the encoded latent space;
however, if the number of mapped features available is small, this would be a very noisy estimate.
In practice, we examined the performance of the autoencoders and cycle consistency outputs with typical validation samples to narrow the search space, although these do not guarantee good mapping performance.
To be fair to the baselines, we tuned our method and Kang method on the same dataset in synthetic data case and the presented results in Figure \ref{fig: Syn2_comp_Radial_GAN_Chimeric_vs_mappedfeatures}-\ref{fig: Syn2_comp_KMF_Chimeric_vs_mapped_features_Onto_case}  are optimal for all methods. 
%In addition, even though the synthetic data generating mechanism that was used to tune the hyperparameters is the same, the samples were independently generated for the final results. 
%In a real deployment, one could subset a single database for experiments to find reasonable hyperparameters as we have done with the ACTFAST dataset above.
In a real example, verification of the proposed mapping with other sources would also give an indication of whether the selected hyperparameters had failed.

We find that there is a substantial tradeoff between the chimeric encoder's ability to \textit{schema match} and its ability to \textit{impute}.
Heuristically, this occurs because \textit{schema matching} using autoencoders requires a substantial amount of compression. 
Only by compression to a low-dimensional latent space using diffuse inputs for each latent variable are we able to closely align the semantics of the two databases.
Otherwise, one can easily imagine a latent space that ``separates'' into a term for the known-mapped features and unmapped features.
However, with mapping mostly accomplished, learning to impute and transform benefits from a lower level of compression and regularization.
Fortunately, the simple initialization with KMF seems to get most of the ``easy'' cases and so we can let the chimeric encoder optimize for imputation and transformation.

We illustrated our proposal with numeric and binary features, or categorical features easily transformed to those types.
Many real databases will have more complex data types, which will require modification of this approach.
For example, lists of diagnostic codes using the ICD10, ICD9, or SNOMED CT ontologies would need an appropriate embedding applied.
Because these structured data are recognizable by their formats and have fully-developed equivalence maps, we assume that in practice they would be ``known-mapped'' features if present in both databases.
Applying the same (pre-trained) embedding to both databases would create comparable numeric columns.
If structured data was present in only one database, the ability to map features to the existing ontology would depend on the embedding chosen and creating a suitable loss function for the hierarchy of entities in that ontology.
For example, the AHRQ CCS system creates a moderate number of categorical variables representing less fine-grained detail in the ICD9 and ICD10 ontologies, and these would be suitable for mapping.
Mapping features containing text documents, for example, determining that a pair of text features were both ``discharge summaries,'' is likely better handled by techniques specific to that task.
However, if text columns are mapped, they can be embedded to a numeric space and treated as known-mapped features in our proposal.

Missing data creates both logistic and conceptual difficulties for transformation between databases.
The correct transformation of \textit{informatively missing} but \textit{easily imputed} data could be either ``missing'' or the unobserved value.
For example, many patients do not have invasive arterial blood gas measurements, because they are measured only under special clinical concerns, but they do have non-invasive measurements which are closely related.
Whether or not the two ought to be mapped will depend on the goals of the project.
To simplify our presentation, we assumed an imputation model was already available to fill in sporadic missing data, but it is straightforward to combine masking strategies used in imputing autoencoders \cite{pereira_reviewing_2020} with our pipeline for a more integrated algorithm.

Finally, we observed the optimization problem behind the chimeric encoder to be a difficult one.
The achieved loss functions were often far from the optimum in cases where it performed poorly, and the optimization was very dependent on batch size and other learning hyperparameters.
This behaviour suggests that further experimentation with initialization and optimization strategies could meaningfully improve its performance.
As these details are likely to be problem- and architecture-specific, we have not exhaustively explored them.

\section{Conclusion}
We study the problem of schema matching across databases using Electronic Health Records (EHRs) from different hospitals or from different time points as a motivating example. 
We address this problem in two stages: first, we find many easy or exact matches using a pairwise correlation-based fingerprint method. Second, we improve on those matches and transform or impute features without a direct match using deep learning. 
Our proposed method relies on the dependencies between a few known mapped features and the rest of the unmapped features. 

% \subsection{The ``Teaser Figure''}

% A ``teaser figure'' is an image, or set of images in one figure, that
% are placed after all author and affiliation information, and before
% the body of the article, spanning the page. If you wish to have such a
% figure in your article, place the command immediately before the
% \verb|\maketitle| command:
% \begin{verbatim}
%   \begin{teaserfigure}
%     \includegraphics[width=\textwidth]{sampleteaser}
%     \caption{figure caption}
%     \Description{figure description}
%   \end{teaserfigure}
% \end{verbatim}

% \begin{acks}
% \end{acks}

\bibliographystyle{plain}
\bibliography{SchemaMatching.bib}

\appendix

\setcounter{table}{0}
\renewcommand\thetable{\Alph{section}.\arabic{table}}

\setcounter{figure}{0}
\renewcommand\thefigure{\Alph{section}.\arabic{figure}}

\section{Appendix} 

\subsection{Code}\label{codeappendix}
Code for the algorithm and synthetic data experiments are available at  \href{https://github.com/sandhyat/KMFChimericE_SchMatch}{github.com/sandhyat/KMFChimericE\_SchMatch}. 
Neither MIMIC nor ACTFAST is available without restrictions, and so the raw data is not included.
Processing code for MIMIC and ACTFAST are included.

\subsection{Synthetic Data} \label{synth_details}
We generate covariance matrices by $\mathbf{W}\mathbf{W}^T + D$ where $\mathbf{W}$ is a spherical Gaussian matrix of size $(n,k)$ and $D$ is a diagonal matrix with $n$ integer values randomly drawn from the interval $[1,20]$. 
Here, $k$ denotes the underlying factor dimension of the generated data. For the 2-cluster Gaussian case, we use common covariance matrix and mean vectors sampled from a uniform distribution. For each data generating mechanism, we sample $5$ datasets of $10,000$ examples.

For the case where the two databases have equal number of columns and onto mapping where smaller set of column is always a subset of the larger column set, we demonstrate the performance of proposed algorithm on samples from following two data generating distributions.
\begin{enumerate} \label{gaussianGen}
    \item \textbf{2-cluster Gaussian simulated (20-D) }: Mixture of two $20$-dim Gaussians with mean vectors randomly sampled from $[10,20]$ separately. The true factor dimension is taken to be $k=10$.
    \item \textbf{Multivariate Gaussian simulated (20-D)}: $20$-dim Gaussian with covariance as above.
\end{enumerate}
% We also use \textbf{2-cluster Gaussian simulated (20-D)} and \textbf{SD2} for the case when in addition to the columns being permuted in the two databases, some columns have affine transformations or nonlinear transformation mapping.
For the partial map example, where the number of columns in two databases are not equal and there is an incomplete overlap between the feature sets, we use the following data generating mechanism:
\begin{enumerate}
    % \item \textbf{SD5}: $50$-dim mean zero Gaussian covariance matrix generated in a similar way as for \textbf{SD2}. \textit{And columns sampled at random ...}
    \item \textbf{Multivariate Gaussian simulated (50-D)}: $50$-dim Gaussian with covariance matrix generated as above.
\end{enumerate}

To account for randomness while partitioning the dataset and selecting the set of mapped variables, for a fixed number of mapped features, we repeat the experiment $n_t$ times with different samples of pre-mapped features. Also, to account for the randomness in permutation or the features that are added/removed, for a fixed set of mapped features, we repeat the experiment $n_p$ times. So, for a given number of pre-mapped features, we repeat the experiment $n_t*n_p$ times.

\subsection{ACTFAST data details} \label{app: real_Data_details}
Feature names and their types for ACTFAST data that were used in this study are presented in Table \ref{tab: real_Data_feature_names}. The pre-mapped feature set was chosen from the categorical features that were further one-hot encoded. Table \ref{tab:smart_data_example} shows how the metadata is not always useful for mapping two databases. 
% In Table \ref{tab:hypertension_icd10_codes}, we present the ICD-10 codes for the feature `hypertension'. 

\begin{table}[h!]
{\scriptsize
\centering
\begin{tabular}{|c|c|c|c|}
\hline
\textbf{Categorical features} & \multicolumn{3}{c|}{\textbf{Binary features}}                   \\ \hline
Anesthesia Type               & HTN                    & CAD                & CAD prior MI      \\ \hline
Valvular disease              & CHF Diastolic function & AFIB               & PPM ICD           \\ \hline
CPAP usage               & CV TIA Stroke          & PAD                & DVT               \\ \hline
ASA                    & CKD                    & Outpatient insulin & Dialysis history  \\ \hline
PAP Type                           & PHTN                   & COPD               & Asthma            \\ \hline
Surgery type                      & OSA                    & Cirrhosis          & Cancer\_hx        \\ \hline
Functional Capacity                  & GERD                   & Anemia             & Coombs pos        \\ \hline
Sex           & Dementia               & Smokingever        & Stopbang observed \\ \hline
Race                          & Stopbang pressure      & Stopbang snore     & Stopbang tired    \\ \hline
                          & CHF                    & DM                 & PE                \\ \hline
                              &                        & Emergency          &                   \\ \hline
\end{tabular}}
\caption{Feature names for ACTFAST data used in Section \ref{sec: real_data_experiments}}
\label{tab: real_Data_feature_names}
\end{table}

% \begin{minipage}{.56\textwidth}
{\small
    \begin{table}[H]
    \centering
    \begin{tabular}{|c|c|}
    \hline
    \textbf{Epic feature name} & \textbf{Contextually derived name} \\ \hline
    BW\#608 & Hypertension \\ \hline
    EPIC\#55845 & Diagnosis year \\ \hline
    EPIC\#31000009721 & Typical SBP \\ \hline
    EPIC\#31000040796 & Typical DBP \\ \hline
    EPIC\#5266 & Hyperlipidemia \\ \hline
    EPIC\#HPI0118 & CAD \\ \hline
    EPIC\#20764 & CCS angina class \\ \hline
    BW\#164 & No revasc but \textgreater 50\% stenotic coronary \\ \hline
    EPIC\#10072 & MI \\ \hline
    BW\#165 & Number of MIs \\ \hline
    EPIC\#62854 & Date of last MI \\ \hline
    EPIC\#3710 & History of CABG \\ \hline
    EPIC\#62856 & Prior CABG date \\ \hline
    \end{tabular}
    \caption{Table showing example features where metadata (feature names) can not be used by language model algorithms for schema matching.}
    \label{tab:smart_data_example}
    \end{table}}
% \end{minipage}
% \begin{minipage}{.4\textwidth}
% {\small
%     \begin{table}[H]
%     \centering
%     \begin{tabular}{|cc|}
%     \hline
%     \multicolumn{2}{|c|}{\textbf{ICD-10 codes for hypertension}} \\ \hline
%     \multicolumn{1}{|c|}{'I10'} & 'I150' \\ \hline
%     \multicolumn{1}{|c|}{'I110'} & 'I151' \\ \hline
%     \multicolumn{1}{|c|}{'I119'} & 'I152' \\ \hline
%     \multicolumn{1}{|c|}{'I120'} & 'I158' \\ \hline
%     \multicolumn{1}{|c|}{'I129'} & 'I159' \\ \hline
%     \multicolumn{1}{|c|}{'I130'} & 'I160' \\ \hline
%     \multicolumn{1}{|c|}{'I1310'} & 'I161' \\ \hline
%     \multicolumn{1}{|c|}{'I1311'} & 'I169' \\ \hline
%     \multicolumn{1}{|c|}{'I132'} & 'I674' \\ \hline
%     \end{tabular}
%     \caption{ICD-10 codes for hypertension that did not always belong in the hierarchical structure available due to different causes for hypertension.}
%     \label{tab:hypertension_icd10_codes}
%     \end{table}}
% \end{minipage}

% \subsection{Flow charts for experimental procedure} \label{app: flow_chart_exp_procedure}

\subsection{Null case result verification on synthetic data} \label{app: null_case}
In this section, we present the null behaviour of our model. To achieve this, we generated a 20-dimensional Gaussian dataset with covariance as identity matrix. As in this case, there is no correlation among the features, choosing any number of mapped features cannot provide the information about the others and hence can not be used by the decoder of other AE. 
%We set up the experiment in a similar way as described in Section \ref{subsubsec: Exp_org_syn}. 
The results from this experiment are illustrated in Figure \ref{fig: Syn3_Chimeric_vs_mappedfeatures}. Clearly, even when the number of features are as high as 10, the fraction of mistakes is still close to $1$.

\subsection{Effect of latent representation size} \label{app: L_dim_comp}
\begin{figure*}[htbp!]
    \centering
    \begin{subfigure}[b]{0.478\textwidth}
        \centering
        \includegraphics[width=1.0\textwidth]{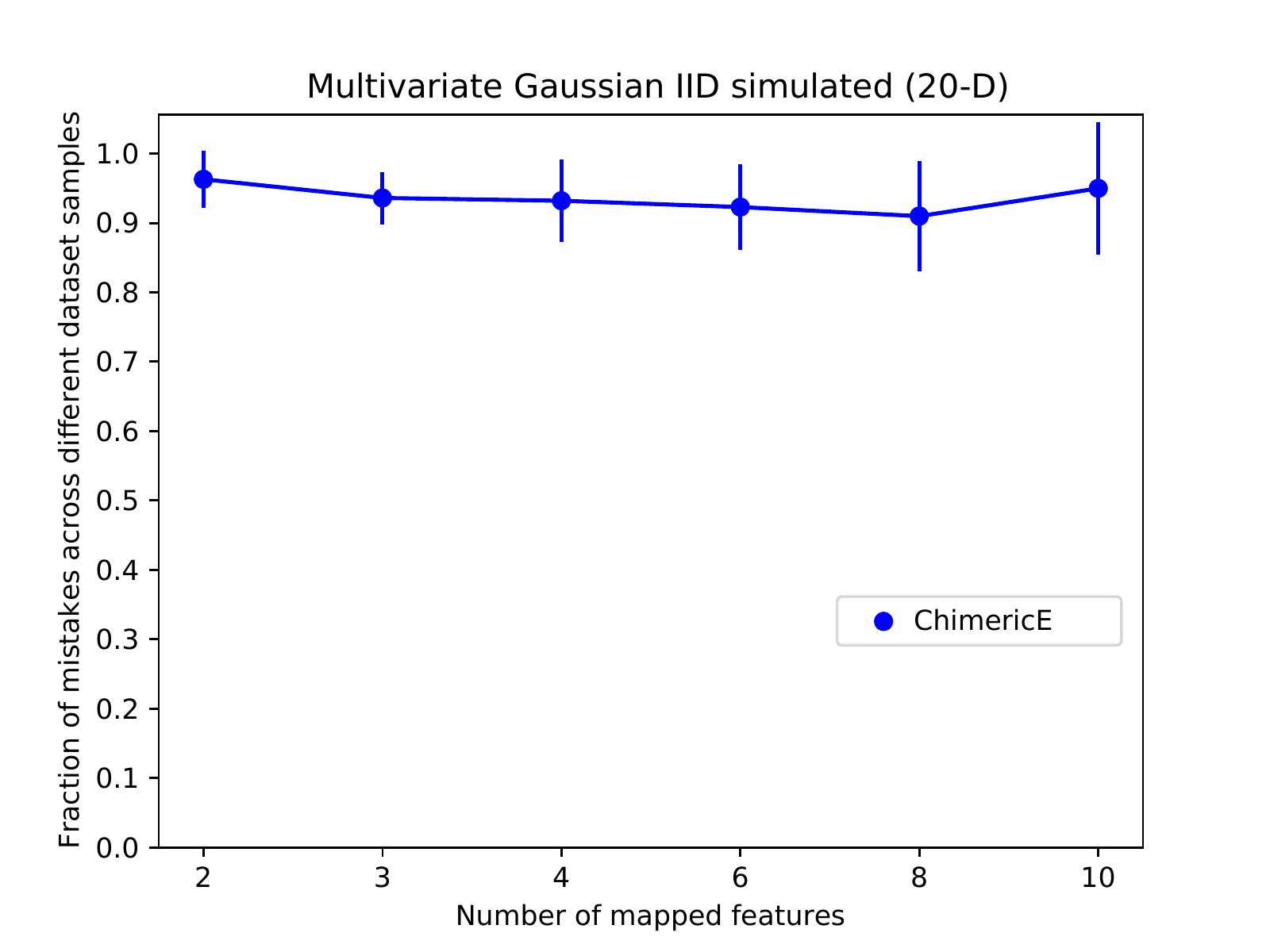}
        \caption{\footnotesize{}}
        % {}    
        \label{fig: Syn3_Chimeric_vs_mappedfeatures}
    \end{subfigure}
    \hfill
    \begin{subfigure}[b]{0.478\textwidth}  
        \centering 
        \includegraphics[width=1.0\textwidth]{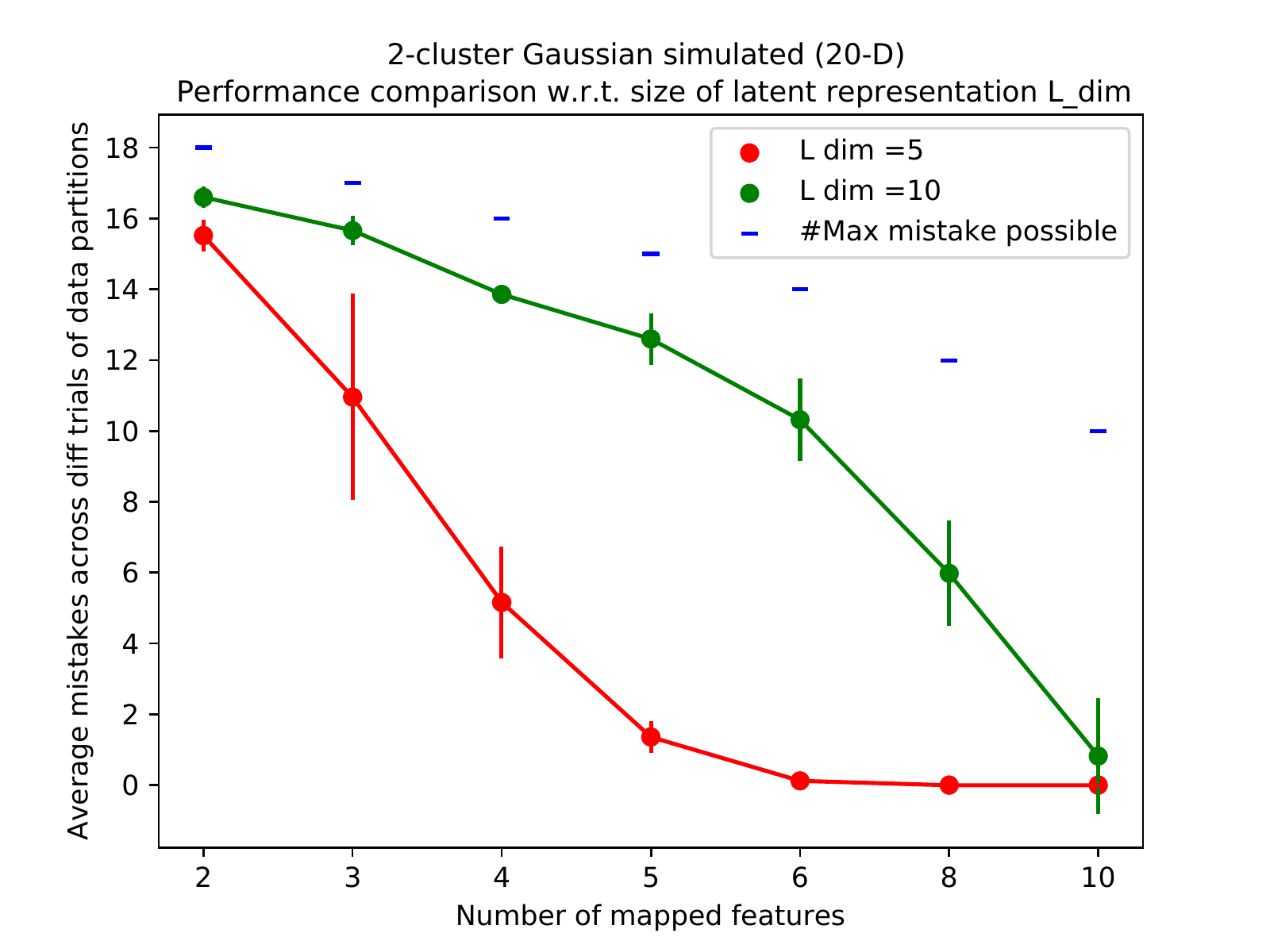}
        \caption{\footnotesize{ }}
        % {}       
        \label{fig: SD1_compar_L}
    \end{subfigure}
        \caption{(a) Performance of chimeric AE on a synthetic dataset \textbf{Multivariate Gaussian IID simulated (20-D)} in which there is no inter-feature dependence within the feature set. As expected, the method performs badly in the absence of any correlations between the mapped and unmapped features. (b) chimeric encoder's performance variation on \textbf{2-cluster Gaussian simulated (20-D) } w.r.t. the size (L dim) of the latent representation; higher compression (small L dim) leads to good  matching performance.}
\end{figure*}

In Figure \ref{fig: SD1_compar_L}, we consider \textbf{2-cluster Gaussian simulated (20-D) } dataset to demonstrate the observation that smaller latent dimension is beneficial as far as matching features is concerned. 

\subsection{Additional results for learning some non-linear transformations and partial map experiments with step down procedure} \label{app: no-match_FDR_examples}

\begin{figure*}[htbp!]

    \centering
    \begin{subfigure}[b]{0.478\textwidth}
        \centering
        \includegraphics[width=1.0\textwidth]{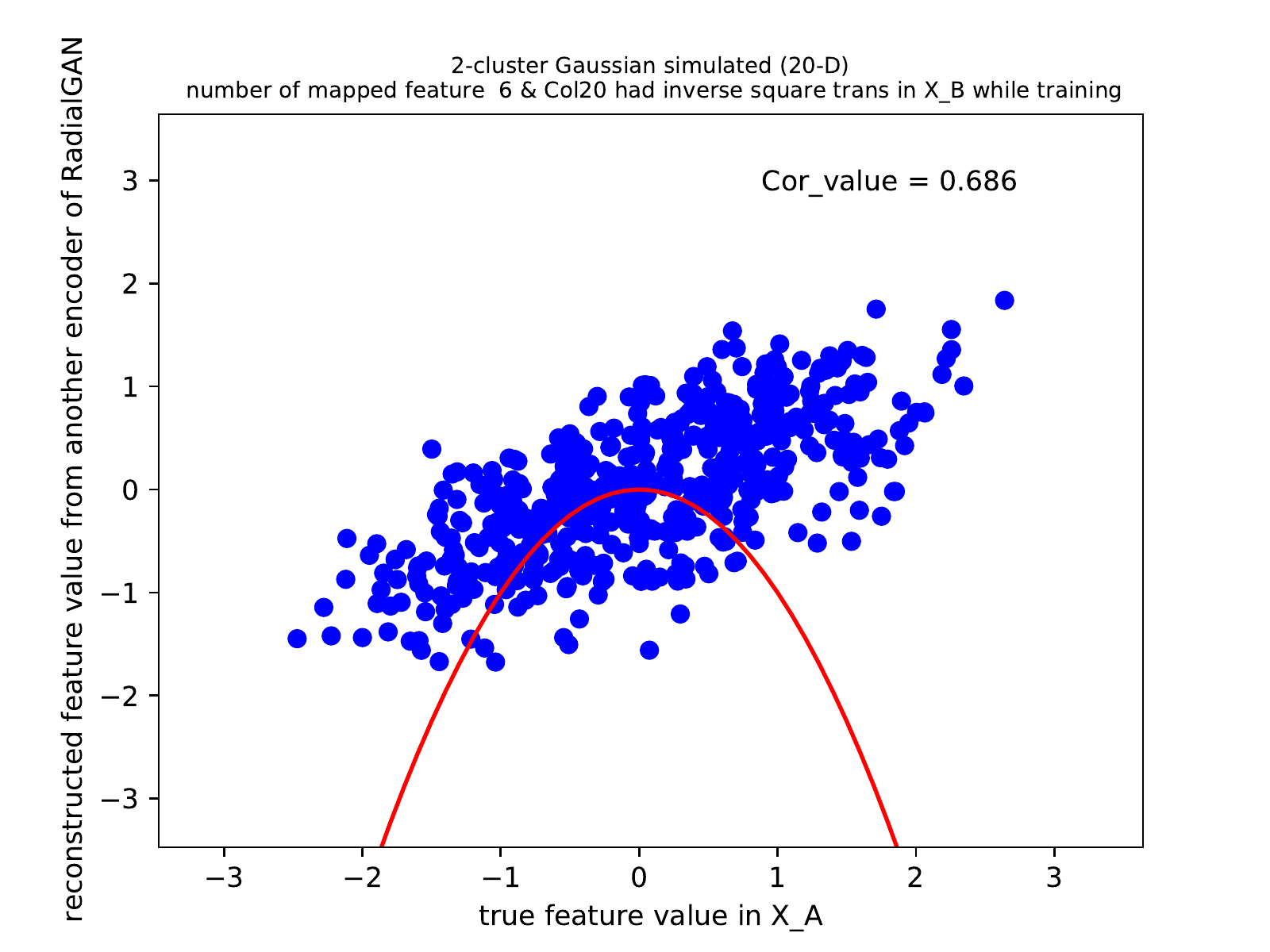}
        \caption{\footnotesize{}}
        % {}    
        \label{fig: RG_not_learning_inv_trans}
    \end{subfigure}
    % \begin{subfigure}[b]{0.478\textwidth}
    %     \centering
    %     \includegraphics[width=1.0\textwidth]{Apdx_figures/Tidal_volumume_set.pdf}
    %     \caption{\footnotesize{}}
    %     % {}    
    %     \label{fig: MIMIC_dist_change_Tidal_volume}
    % \end{subfigure}
    \vskip\baselineskip

    \begin{subfigure}[b]{0.478\textwidth}
        \centering
        \includegraphics[width=1.0\textwidth]{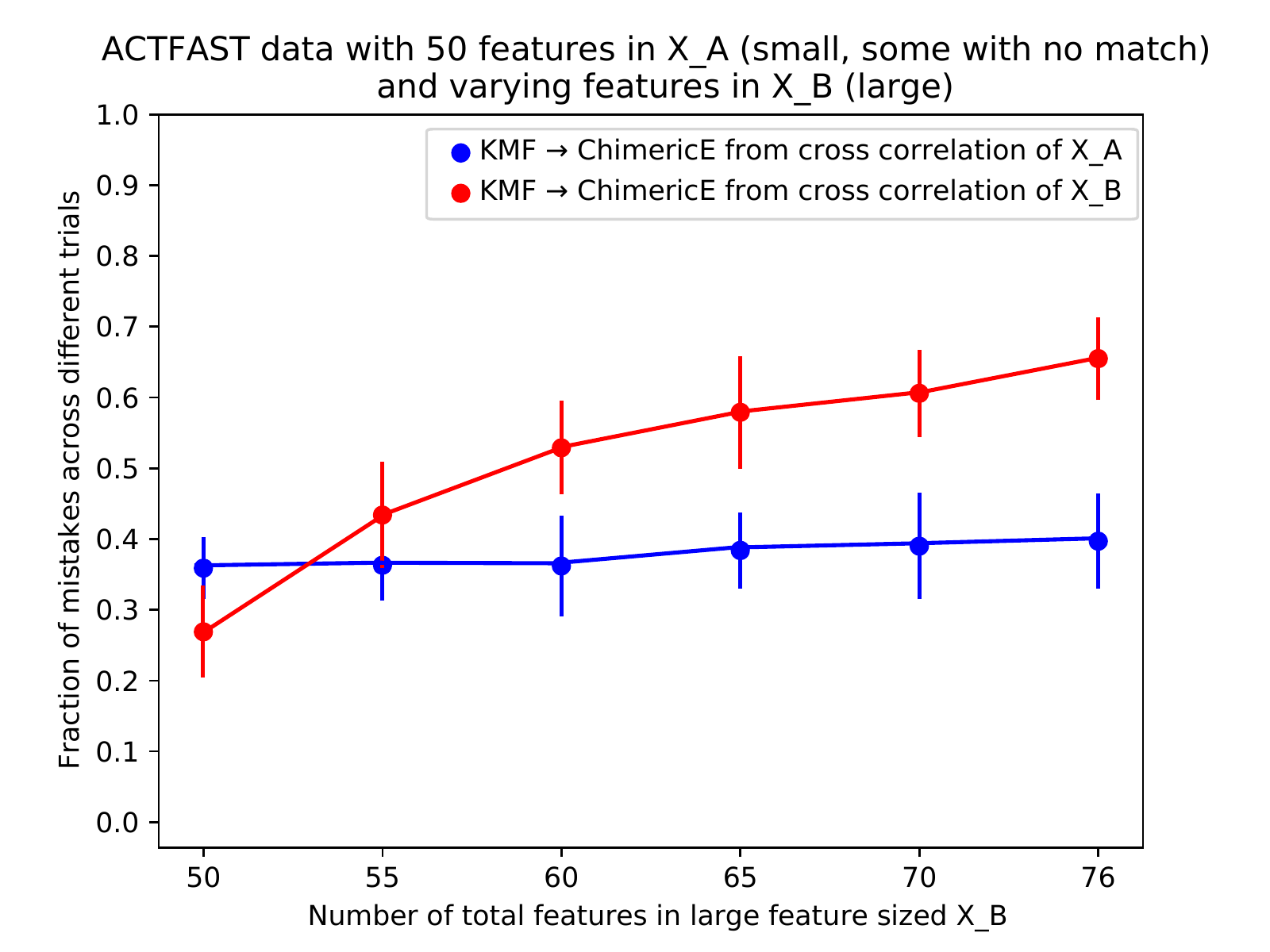}
        \caption{\footnotesize{}}
        % {}    
        \label{fig: REal_Data_partial_mapping_5_extra_two_Stage_from_CCx1_vsCCx2}
    \end{subfigure}
    \begin{subfigure}[b]{0.478\textwidth}
        \centering
        \includegraphics[width=1.0\textwidth]{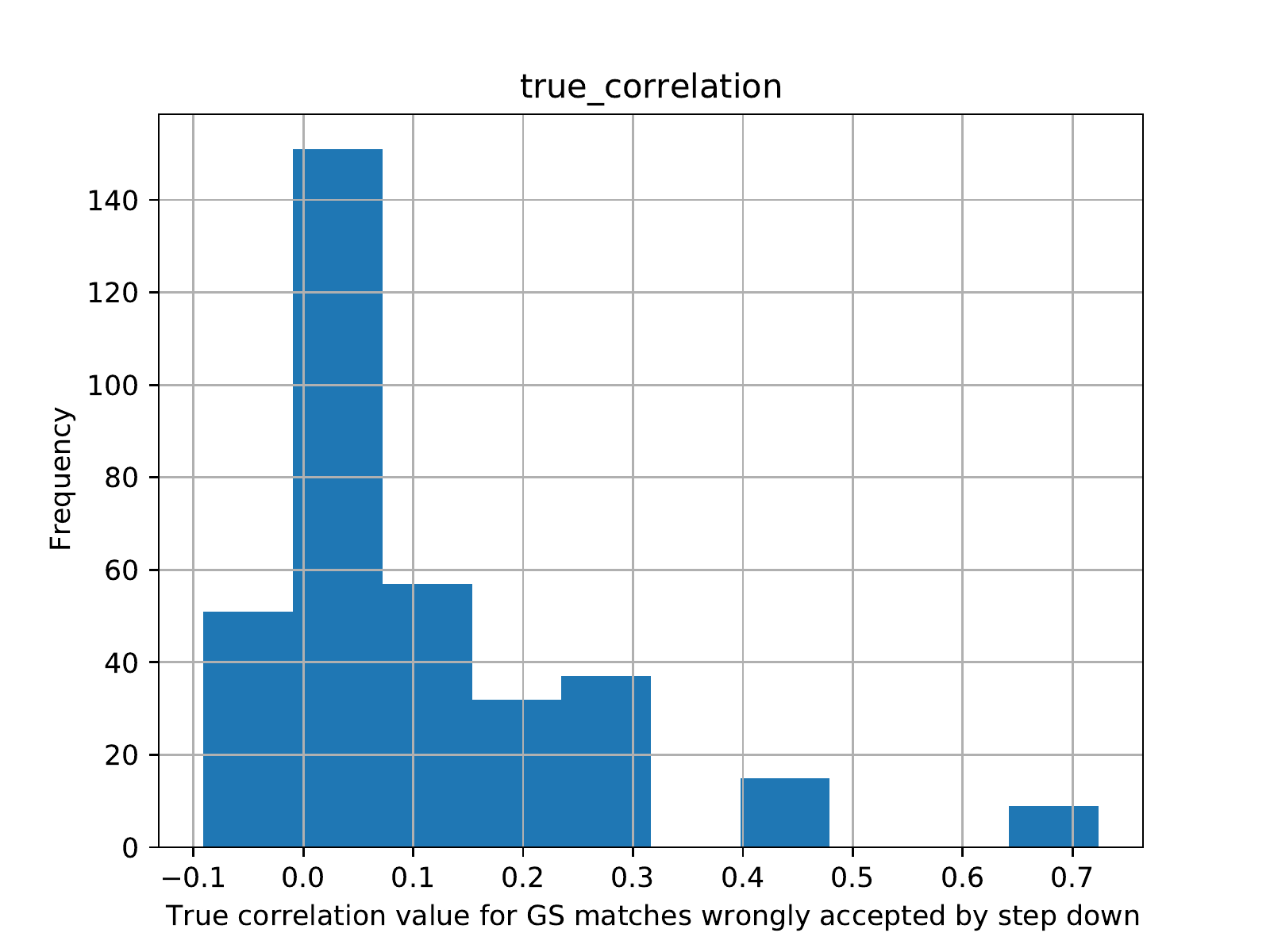}
        \caption{\footnotesize{}}
        % {}    
        \label{fig: Real_data_no_true_match_GS_matches_correlation_SD_acceptes_fromX1}
    \end{subfigure}
    \caption{ \textbf{2-cluster Gaussian simulated (20-D)} (a) RadialGAN transformed feature data; actual transformation to be learnt is an inverse square transformation which RadialGAN fails to learn for the same settings as in Figure \ref{fig: Syn1_Chimeric_vs_mappedfeatures_Invsq_transformed_Recons_Col20}. 
    % \textbf{MIMIC-III dataset} (b) Example to show that there could be distribution shift for same feature in two databases to match; in this case, the two databases correspond to two different EHR systems in different times.
    \textbf{ACTFAST data} (b) Example to show the difference in the performance in case of onto or partial map case when two different proposing directions are used in GS algorithm. (c) Histogram of true correlation between `no-match' features and their proposed match using GS algorithm. These matches were wrongly accepted as correct by step down procedure as they act as surrogate matches due to high correlation. }
\end{figure*}

We first present a scatter plot that demonstrates the ability of RadialGAN to learn an inverse square transformation. As can be seen in Figure \ref{fig: RG_not_learning_inv_trans}, RadialGAN is not even able to learn the sign correctly, whereas chimeric encoder could learn an inverted parabola as seen in Figure \ref{fig: Syn1_Chimeric_vs_mappedfeatures_Invsq_transformed_Recons_Col20}.

% We use the MIMIC-III dataset to demonstrate the situation when there could be some features with distribution shift between two databases. For example, as seen in Figure \ref{fig: MIMIC_dist_change_Tidal_volume}, we observe a decrease in the value of feature `Tidal volume (set)' in MV era. This could be because of the studies that started presenting evidence that lower tidal volume is better \cite{acute2000ventilation}. Such kind of examples are common and hence pose a difficulty for schema matching methods that rely only on feature distribution statistics.

Moving to the incomplete feature overlap case, we provide an example in Figure \ref{fig: REal_Data_partial_mapping_5_extra_two_Stage_from_CCx1_vsCCx2} that shows the difference between the fraction of mistakes obtained from the two different direction of proposals as the size of one of the databases is increased. As explained earlier this is due to the inability of chimeric encoder (that generates $z_A$) to reconstruct the latent representation (of $x_B$) containing information of larger number of features than $x_A$ and hence the performance worsens (red curve ) as the number of features in $x_B$ increase. This phenomenon is not true vice versa as the latent representation of $x_A$ has more flexibility when the chimeric encoder transforms it to $z_B$ and hence the performance is invariant (blue curve) to increase in feature size.

Next we present the example where the match for an originally `no-match' feature is a surrogate and we demonstrate it by providing the density of the true correlations on the GS matches for `no-match' features that were declared significant by step down (mistakes) in Figure \ref{fig: Real_data_no_true_match_GS_matches_correlation_SD_acceptes_fromX1}. Some examples of such GS matched feature pairs used to obtain the above mentioned density along with their frequency and the corresponding correlation are reported in Table \ref{tab: no-match_mistakes_correlation}.

\begin{table}[!h]
{\scriptsize
\begin{tabular}{|c|c|c|c|}
\hline
\textbf{match by GS for no-match features}  & \textbf{frequency} & \textbf{true\_correlation} & \textbf{estimated\_cross\_corr} \\ \hline
( HTN , StopBang\_Pressure )                & 9                  & 0.723                      & 0.947                           \\ \hline
( Dialysis\_History , CKD )                 & 1                  & 0.426                      & 0.175                           \\ \hline
( CKD , Dialysis\_History )                 & 4                  & 0.426                      & 0.343                           \\ \hline
( PPM\_ICD , CHF )                          & 9                  & 0.415                      & 0.419                           \\ \hline
( OSA , CPAP.Usage\_1 )                     & 1                  & 0.408                      & 0.150                           \\ \hline
( OSA , StopBang\_Observed )                & 14                 & 0.288                      & 0.164                           \\ \hline
( PAP\_Type\_1 , CAD )                      & 7                  & 0.268                      & 0.361                           \\ \hline
( OSA , StopBang\_Snore )                   & 1                  & 0.248                      & 0.161                           \\ \hline
( CKD , CHF\_Diastolic\_Function )          & 10                 & 0.242                      & 0.357                           \\ \hline
( PAP\_Type\_1 , PHTN )                     & 5                  & 0.235                      & 0.324                           \\ \hline
( CKD , CAD\_PRIORMI )                      & 3                  & 0.230                      & 0.468                           \\ \hline
( PAP\_Type\_1 , CHF\_Diastolic\_Function ) & 1                  & 0.222                      & 0.300                           \\ \hline
( OSA , CPAP.Usage\_2 )                     & 1                  & 0.208                      & 0.152                           \\ \hline
( OSA , HTN )                               & 1                  & 0.205                      & 0.185                           \\ \hline
\end{tabular}}
\caption{List of `no-match' features and their corresponding match from chimeric encoder across different permutations and different trials. The high value of true correlation implies there are features that have surrogate matches that FDR step down procedure is not able to identify as mistakes. }
\label{tab: no-match_mistakes_correlation}
\end{table}

\subsection{Model hyperparameters} \label{subsubsec: Model_arc_syn}
In all the experiments, we chose both the encoders and decoders to be multi-layer perceptrons with two hidden layers each. For synthetic and ACTFAST datasets, hidden units for first and second layer is 80 and 40 respectively for encoder and vice versa for the decoder. 
For MIMIC experiments, the hidden layer neurons were chosen to be 120 and 70. 
For the optimizer, we used Adam optimizer with weight decay of $10^{-5}$ along with a learning rate scheduler. The initial learning rate of the scheduler for synthetic and ACTFAST experiments is fixed at $10^{-2}$, but it was tuned in the MIMIC experiments.

For the synthetic data experiments, we used \textit{tanh} activation and dropout after second hidden layer of encoder and after first hidden layer of the decoder. 
% The output layer of encoder (latent space) was tuned over $l \in \{5,8,10,12\}$.
We used batch size of $64$ for both $x^A$ and $x^B$. 
% For the optimizer, we used Adam optimizer with weight decay of $10^{-5}$ along with a learning rate scheduler that starts at learning rate value of $10^{-2}$. 
Number of epochs for all synthetic data experiments was kept at $40$ for chimeric AE. 
%We weight the three losses viz., direct reconstruction loss, cross reconstruction loss (Eq. \ref{crossloss}) and cycle consistency loss (Eq. \ref{cycleloss}) differently.
% by choosing the weights from $ \{0.5,0.8,0.9, 1.0, 1.1, 1.2 \}$.  
We encouraged orthogonalization of the latent space with a regularization loss $\Vert f^i(x^i)^Tf^i(x^i) - \mathbb{I}_l\Vert_2$ evaluated within batches with weighing parameter value $w_o = 0.01$. For the binarized synthetic data experiment in Figure \ref{fig: Syn1_Chimeric_vs_mappedfeatures_binarized}, we use the same settings as above except for no tanh activation within the network and add sigmoid activation at the end of the encoders.
Hyperparameter optimization was done on one of the synthetic datasets (\textbf{Multivariate Gaussian simulated (20-D)}), and the set with best matching performance was chosen and used in all other Gaussian synthetic dataset examples. % except for binarized data which required separate tuning.
We compared dropout rate $p \in \{0.5,0.6,0.7 \}$, encoder dimension $l \in \{5,8,10,12\}$, and the weights for different loss terms from $ \{0.5,0.8,0.9, 1.0, 1.1, 1.2 \}$ in cross-validation. 
We set the FDR value in the mapping step at $0.05$.

For MIMIC data experiments, we used \textit{tanh} activation and dropout after the second hidden layer of the encoder and after the first hidden layer of the decoder. 
For other hyperparameters, we performed a grid search as follows: dropout rate $p \in \{0.4,0.5,0.6,0.7, 0.8\}$, latent representation size $l \in \{20,30,40\}$, batch size $\in \{32, 64\}$, learning rate $\in \{10^{-2}, 10^{-3}\}$ and the weights for different loss terms $\in \{ 0.5,0.7,0.8, 1.0, 1.1, 1.2, 1.4 \}$. 
Number of epochs was set at 50 for chimeric encoders.

For the ACTFAST data experiments, we add sigmoid activation at the end of the encoders. We used \textit{ReLu} activation (only in onto and partial map case) and dropout with dropout rate $p \in \{0.5,0.6,0.7 \}$ after second hidden layer of encoder and after first hidden layer of the decoder. We used batch size of $32$ for both $x^A$ and $x^B$. The output layer of encoder (latent space) was tuned over $l \in \{10,20,30,40\}$. Number of epochs for all data experiments were set at $50$ for chimeric encoders. Other settings remain same as for the synthetic experiments described above. 
Hyperparameter tuning was done separately for permutation case and onto map case with matching performance as the criterion.
The best set from onto map experiment was also used in partial map training except for latent space dimension which was $20$ for onto and $10$ for partial map. 
In all 3 experiments for ACTFAST dataset, we set number of trials $n_t=4$ and number of permutations within a trial as $n_p=3$. 

For the Kang method we optimize Euclidean distance for the permutation map case and Normal distance for the onto and partial map case with $\alpha $ tuned in the set $\{1,1.2,1.4,1.6,1.8,2,,4,5,10\}$. 3000 iterations were used in the Kang method search for Euclidean distance and 5000 for Normal distance. 
To choose the best value of $\alpha$, we ran the experiments over 2 trials and 2 permutations and then for the final experiment, use the value that has the highest F1 score. For the number of iterations, we gradually increased the number of iterations, until the average performance saturated and used the corresponding number in our experiments.

For RadialGAN, we use the same architecture settings and hyperparameter tuning range as suggested by the authors and run the experiment for 100 epochs (no changes in performance were noted out to 1000 epochs in initial experiments).

\end{document}